\documentclass[12pt]{article}
\usepackage{a4wide}
\title{\textbf{Job Prospects and Labour Mobility in China}\thanks{\scriptsize{Acknowledgements: This research received funding from the European Union's Horizon 2020 research and innovation program under the Marie Sklodowska-Curie grant agreement No. 838534. The authors would like to thank Vicente Royuela and session participants at the Chinese Economists Society Annual Conference 2021, the 9th International Conference on Applied Research in Economics and Finance, the 13th Chinese Economic Association and 32th CEA UK Annual Conference, the 46th Symposium of the Spanish Economic Association, the 15th RGS Doctoral Conference in Economics, and the 1st International Workshop on the Chinese Development Model for helpful discussions and comments on the research. Any shortcomings are our own.}}  \textsuperscript{, }\thanks{\scriptsize{Declaration of Interest: The authors declare that they have no known competing financial interests or personal relationships that could have appeared to influence the work reported in this paper.}}}
\author{Huaxin Wang-Lu\thanks{\scriptsize{Corresponding author. Department of Quantitative Methods, IQS School of Management, Universitat Ramon Llull, Via Augusta 390, 08017 Barcelona, Spain. Tel.: +34 637 170 126. E-mail: \color{blue}{huaxin.wanglu@iqs.url.edu.}}} \and Octasiano Miguel Valerio Mendoza\thanks{\scriptsize{Department of Quantitative Methods, IQS School of Management, Universitat Ramon Llull. E-mail: \color{blue}{octasiano.valerio@iqs.url.edu.}}}}
\date{July 2022}
\usepackage{microtype}
\usepackage[margin=2.5cm]{geometry}
\usepackage{sectsty} 
\usepackage{mathpazo}
\usepackage{antpolt}
\usepackage[T1]{fontenc}
\usepackage{libertine}
\subsectionfont{\itshape}
\subsubsectionfont{\normalfont\itshape}
\usepackage[section]{placeins} 
\usepackage{anyfontsize}
\usepackage[euler]{textgreek}
\usepackage{amsmath}
\usepackage{lipsum} 
\usepackage{wasysym} 
\usepackage[nodisplayskipstretch]{setspace} 
\usepackage{graphicx} 
\usepackage{stfloats} 
\usepackage{subcaption} 
\usepackage{longtable} 
\usepackage{supertabular}
\usepackage{multirow} 
\usepackage{ragged2e,array,booktabs} 
\usepackage{threeparttablex}
\usepackage{rotating} 
\usepackage{pdflscape}
\usepackage[table]{xcolor} 
\usepackage{mwe}
\usepackage{booktabs} 
\usepackage[normalem]{ulem} 
\usepackage{caption} 
\usepackage{threeparttable}  
\usepackage[style=authoryear-icomp,doi=true, giveninits, isbn=false,url=false,eprint=false,backend=biber,autocite=inline, labeldateparts=true, uniquename=false, uniquelist=false, ibidtracker=false, pagetracker=true, sortcites=true]{biblatex}

\DeclareNameAlias{sortname}{family-given}

\DeclareNameWrapperFormat{labelname:poss}{#1's}

\newrobustcmd*{\posscitealias}{%
  \AtNextCite{%
    \DeclareNameWrapperAlias{labelname}{labelname:poss}}}

\newrobustcmd*{\posscite}{%
  \posscitealias
  \textcite}

\newrobustcmd*{\Posscite}{\bibsentence\posscite}

\newrobustcmd*{\posscites}{%
  \posscitealias
  \textcites}

\usepackage[hypertexnames=false,citecolor=blue,colorlinks=true,linkcolor=blue,urlcolor=black,bookmarksopen=true]{hyperref} 

\newlength{\mycolwidth}

\usepackage{bookmark}

\graphicspath{{/Users/huaxinwanglu/Library/Mobile Documents/com~apple~CloudDocs/2nd Paper/}}

\DeclareFieldFormat{citehyperref}{%
  \DeclareFieldAlias{bibhyperref}{noformat}
  \bibhyperref{#1}}

\DeclareFieldFormat{textcitehyperref}{%
  \DeclareFieldAlias{bibhyperref}{noformat}
  \bibhyperref{%
    #1%
    \ifbool{cbx:parens}
      {\bibcloseparen\global\boolfalse{cbx:parens}}
      {}}}

\savebibmacro{cite}
\savebibmacro{textcite}

\renewbibmacro*{cite}{%
  \printtext[citehyperref]{%
    \restorebibmacro{cite}%
    \usebibmacro{cite}}}

\renewbibmacro*{textcite}{%
  \ifboolexpr{
    ( not test {\iffieldundef{prenote}} and
      test {\ifnumequal{\value{citecount}}{1}} )
    or
    ( not test {\iffieldundef{postnote}} and
      test {\ifnumequal{\value{citecount}}{\value{citetotal}}} )
  }
    {\DeclareFieldAlias{textcitehyperref}{noformat}}
    {}%
  \printtext[textcitehyperref]{%
    \restorebibmacro{textcite}%
    \usebibmacro{textcite}}}

\newcolumntype{C}[1]{>{\centering\arraybackslash}p{#1}} 

\begin{document}

\maketitle

\begin{abstract}
\singlespacing
China's structural changes have brought new challenges to its regional employment structures, entailing labour redistribution. By now Chinese research on migration decisions with a forward-looking stand and on bilateral longitudinal determinants at the prefecture city level is almost non-existent. This paper investigates the effects of sector-based job prospects on individual migration decisions across prefecture boundaries. To this end, we created a proxy variable for job prospects, compiled a unique quasi-panel of 66,427 individuals from 283 cities during 1997--2017, introduced reference-dependence to the random utility maximisation model of migration in a sequential setting, derived empirical specifications with theoretical micro-foundations, and applied various monadic and dyadic fixed effects to address multilateral resistance to migration. Multilevel logit models and two-step system GMM estimation were adopted for the robustness check. Our primary findings are that a 10\% increase in the ratio of sector-based job prospects in cities of destination to cities of origin raises the probability of migration by 1.281--2.185 percentage points, and the effects tend to be stronger when the scale of the ratio is larger. Having a family migration network causes an increase of approximately 6 percentage points in migratory probabilities. Further, labour migrants are more likely to be male, unmarried, younger, or more educated. Our results suggest that the ongoing industrial reform in China influences labour mobility between cities, providing important insights for regional policymakers to prevent brain drain and to attract relevant talent.

\bigskip
\hspace{-1.4em}\textbf{JEL Classification:} J61; D90; R11; O14; O15

\hspace{-1.4em}\textbf{Keywords:} internal migration; labour mobility; expectations; reference-dependence; China
\end{abstract}

\newpage

\section{Introduction}
\hspace{1.4em}Countries often undergo a drastic change in their employment structures at times of industrial transformation (e.g., \cites{MILNE1988, Edgington1994, Chen2011}), entailing labour redistribution. China is a remarkable case for study under this background. Since the beginning of its reforms and opening-up in late 1978, the GDP-based ratio of the primary, secondary and tertiary sector changed from 31:47:22 in 1979 to 8:38:54 in 2020 \autocite{CSY2021}. This indicates that the tertiary sector almost trebled its contribution to national GDP during this period. In recent years, after being the `world's factory' for decades, the central government launched the `Made in China 2025' and `Dual Circulation' strategies to promote upgrading from a labour-intensive, export-oriented manufacturing economy to a service and consumption-driven one.\par 

\hspace{1.4em}At the same time, China unveiled a proposed revision to the law on vocational education, announcing in 2020 the `Vocational Education Quality Improvement Action Plan' to fill gaps in skilled technicians and to differentiate skillsets across college graduates. On the demand side, regions at different levels of development require different types of skills at different levels of demand. For instance, inland provinces need workers with plant-based skills in response to the relocation of numerous factories from eastern areas taking advantage of lower labour costs \autocites{Qu2012, Qu2013}, whereas the Yangtze River Delta is dedicated to attracting high-tech and managerial professionals for highly-developed manufacturing and service industries \autocite{Wang2020c}. On the supply side, the job prospects for individuals with diverse skills and profiles differ across regions.\par

\hspace{1.4em}In general, the employment situation of workers with low and average skills is dim, because of the widespread use of automated technology, although the service sector is creating new low-tech jobs \autocite{Li2020}. More recently, computerisation has added to jobless growth.\footnote{Jobless growth means that an economy is growing at a reasonable rate without the proportionate creation of new jobs.} \textcite{Frey2017} assert that the majority of routine jobs in manufacturing and a range of sub-sectors within the tertiary sector, such as finance, logistics and administrative support, are at high risk of being replaced by artificial intelligence (AI) technology, a potential concern for China. \textcite{Zhou2020} even suggest that AI technology will replace approximately 278 million jobs in China by 2049. While the creation of non-routine jobs was stagnant during 1990 to 2015, with more than 50\% of employment routinised in 2015 \autocite{Ge2021}.\footnote{According to \textcite{Autor2003}, tasks that rely on well-defined procedures and activities are classified as routine while tasks that require creativity, problem-solving or human interaction are classified as non-routine. Both tasks further subdivide into cognitive or manual skill types.}\par

\hspace{1.4em}The extent of all these (expected) impacts varies across sectors and regions. It is inevitable that job prospects will be subject to change in regional employment structures, stimulating relocation, a plausible reaction to worsening prospects in particular regions. These impacts could go beyond concerns about job opportunities into job mobility and/or security. In other words, labour market conditions tend to be inconsistent with economic growth, at least in the short term \autocites{Prachowny1993, Pehkonen2000, Sahin2015}. These come to our core idea that an employment flourishing industry could provide better job prospects for individuals with migration intentions over those whose job growth shrinks or stagnates, \textit{ceteris paribus}.\footnote{As seen in \hyperref[figb2]{Figure B1}, GDP grew smoothly in Beijing, Shanghai, Guangzhou, and Shenzhen, while employment levels in the four cities fluctuated.}\par

\hspace{1.4em}While, how potential migrants evaluate whether the industrial employment in an area thrives or not? Suppose that all individuals in a city $A$ always earn 2000 Yuan per month. It will be difficult for them to make an assessment of the economy, given no inflation presumed. At the same time, people in city $B$ earn 2000 Yuan per month in 2011 but 1500 Yuan in 2012. The latter population is very likely to believe that the economy is getting worse, at least, more likely than are $A$ residents. Here, as the reference-dependence in prospect theory reveals \autocite{Kahneman1979},\footnote{Reference-dependence is a central principle in prospect theory. It holds that people evaluate information gradually as losses and gains relative to certain reference points or a status quo, rather than as a final state of an absolute outcome.} the deviation ($1500-2000$) from the 2011 income, which is the reference point closest to the 2012 income, generates a negative signal for evaluation and thus, incurs a pain of loss to $B$ inhabitants.\par 

\hspace{1.4em}Let's consider another two cities $C$ and $D$ where people earn, respectively, 2000 and 3000 Yuan per month in 2011 but both 2500 Yuan in 2012. If $B$ inhabitants with migration intentions are aware of the changes, although moving to either $C$ or $D$ will lead to the same increases in their income, they will perceive additional gains from $C$'s positive deviation ($2500-2000$). This is because such an uptrend implies that $C$ has better economic prospects than the others. Kahneman and Tversky highlighted that individuals would even reverse their preferences when identical outcomes are rephrased as gains or losses. As outlined in Figure 1, this analogy manifests why reference-dependence is meaningful in our context.\par

\begin{figure}[!htb]	
\centering
\captionsetup{justification=centering}	
\includegraphics[scale=0.5]{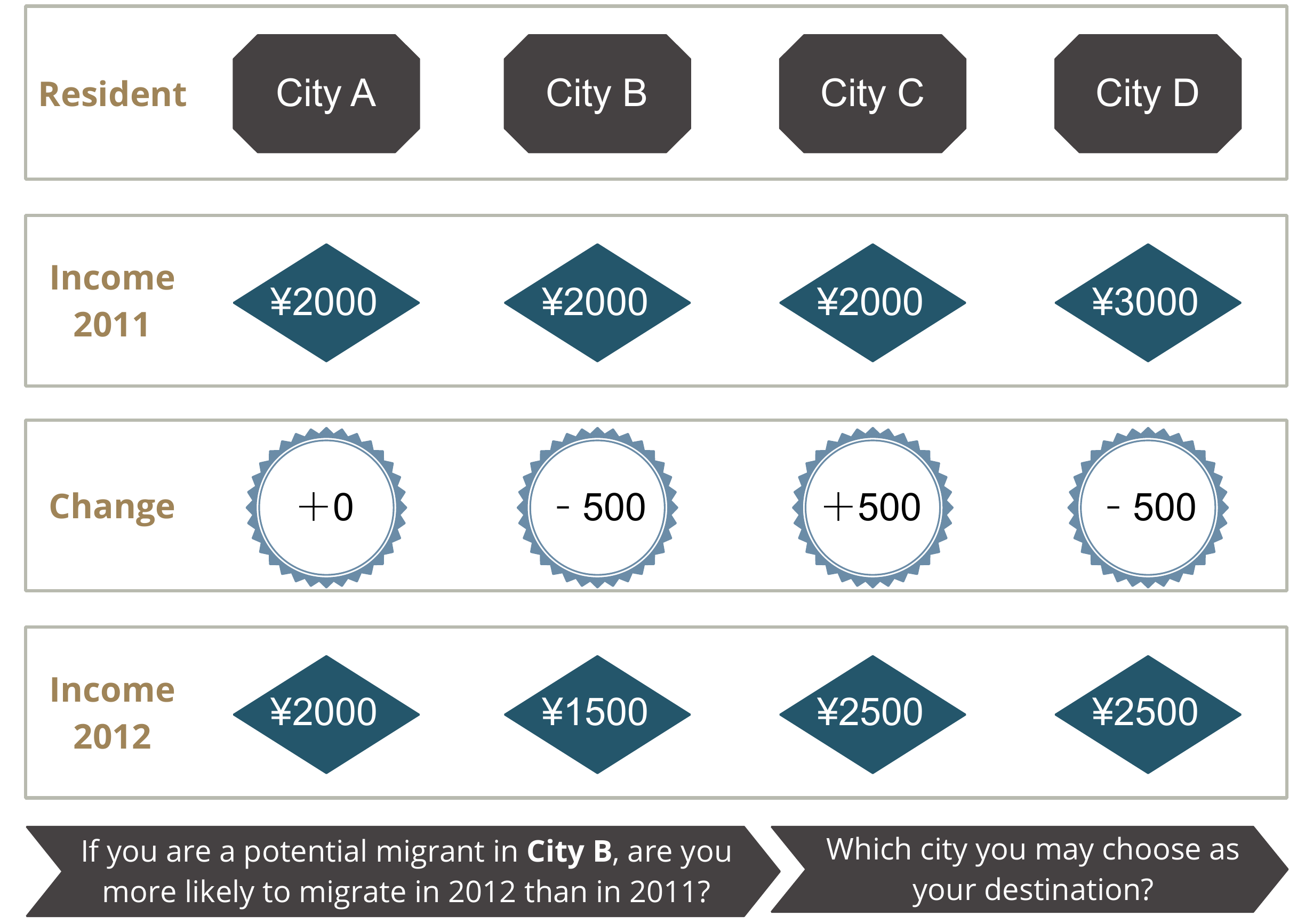}
\caption{\fontsize{10}{12}\selectfont{Analogy for understanding reference-dependence.}}
\end{figure}

\hspace{1.4em}On the other hand, the feature of population redistribution has also been changing. Intra-provincial migration seems to have been more important than inter-provincial migration for both temporary and permanent migrants during this decade \autocites{Zhang2013, Meng2020}. Inter-provincial inequalities have often been associated with the prevalence of inter-provincial migration (e.g., \cites{Kanbur1999, Chan2008, Peng2017}). If this is the case, a co-movement of the popularity of intra-provincial migration and intra-provincial inequalities will arise. Although the cause has not yet been confirmed, Zhejiang showed an abnormally divergent path of the human capital growth from 1985 to 2016 due, in part, to its within-province inequality \autocites{ValerioMendoza2022, Wei2004}. This phenomenon might be a caveat on the imbalance of labour mobility within provinces. Hence, keeping a watchful eye on city-level migration is of greater importance to China than ever before.\footnote{The grant of Hukou is associated with the access to a variety of social programmes provided by the regional government, such as the entitlement to undertake the college entrance examination (Gaokao), to social security and even to house purchase. Prefecture-level cities are usually the main administrative unit in China designated to manage the household registration (Hukou).}\par

\hspace{1.4em}The objective of this paper is to examine the effects of sector-based job prospects on individual migration decisions across prefecture boundaries.\footnote{In China, prefecture cities rank below provinces, the highest non-national level administrative unit, and above counties.} Our contribution to the migration debate is fourfold. Firstly, we contributed to the considerable dearth in Chinese migration literature of visionary migration decisions concerned with optimising future outcomes.\footnote{Migration decisions involving future outcomes across location choices are defined as non-myopic migration (equivalent to `visionary migration' in this study) in a substantial body of migration literature (e.g., \cites{Baldwin2001, Bertoli2016, Gardner2020, Barreda-Tarrazona2021}). In contrast, `Myopia' is a term referring to migration decisions that only depend on the past and/or current situation.} Though a handful of studies touch upon migrants' expectations, such as expected land reallocation (e.g., \cites{Yan2014, Ren2020}), none considers multilateral resistance to migration resulting from the future attractiveness \autocite{Bertoli2013}.\footnote{The term `multilateral resistance to migration' is used to explain influences exerted by the attractiveness of alternative locations on migration rates between any pair of regions.} We understand that this is the first paper to illustrate migration decision-making of visionary labour migrants within China. Secondly, we extended the random utility maximisation (RUM) model of migration by synthesising the virtues of dynamic discrete choice models and reference-dependence to further illustrate the formation of individual expectations.\footnote{It is also worth noting that the model we developed can be applied to other spatial dimensions, not limited to internal migration or China.} To our knowledge, this is also the first time that linkages have been established between the RUM model of migration and prospect theory \autocite{Kahneman1979}.  Thirdly, we derived empirical specifications with theoretical micro-foundations. Lastly, we compiled a unique quasi-panel of 66,427 individuals moving from 283 cities to 279 cities during 1997--2017 and thus, combined city-level bilateral variations with individual and household characteristics, a level of analysis not yet undertaken by existing Chinese migration literature.\footnote{Cross-sectional and/or provincial analyses predominate Chinese literature on migration \autocite{Su2018}. For bilateral migration research, most recently see \textcites{Liu2020, Pu2019, Zhang2018, Cao2018}, and for longitudinal studies, see \textcites{Shi2020, Gao2019, Wu2019, Mu2015}.} While both the new migration patterns and the essential role of prefecture cities in managing Hukou registration signify the importance of understanding city-level push and pull factors.\par

\hspace{1.4em}The remainder of the paper is organised as follows. Section 2 summarises major strands of the relevant literature. Section 3 introduces a reference-dependent RUM model of migration, the econometric modelling and estimation techniques, and the data source and our sample. Section 4 presents the empirical results, including the robustness check. Our discussion and conclusions are given in Section 5.\par

\section{Migration and Expectations}

\subsection{Migration in China}
\hspace{1.4em}Despite the surge of temporary migration since the late 1980s, this phenomenon was not studied empirically until much later, principally due to the lack of data \autocite{Wu1996}. The earlier migration literature relied mainly on macro-data, typically the national population census or the 1\% population sample survey, treating intra-provincial migrants as `non-movers' (e.g., \cites{Lin2004, Poncet2006}). Although more micro-data became available in the late 2000s, e.g., the Longitudinal Survey on Rural-Urban Migration in China (RUMiC) initiated in 2006, and the China Migrants Dynamic Survey (CMDS) begun in 2009, the vast majority of data options remain incompatible with bilateral or inter-city studies.\footnote{The location information is usually provided at the province level, meaning that the sending and receiving cities cannot be identified. Further, although investigators provided city information, most often only cities of destination are known to applied researchers, resulting in that the majority of migration studies focus on analysing the receiving context. Until recently, we have still known less about the reasons that migrants leave their hometowns than why they move to their areas of destination.} More recently, micro-data have been popular for migration analysis as they can help alleviate reverse causality problems and allow researchers to control for individual heterogeneity of relevance. However, longitudinal micro-data are much harder to collect, resulting in only a few migration studies on individual decisions over time. Further, the popularity of the province-level research also stems from the remarkable increase in inter-provincial migrant populations since 1987 \autocite{Liang2001}, the predominant group until 2010, which was then outpaced by intra-provincial migration, as seen in Figure 2. Similar conclusions can be drawn from national statistics on the rural-to-urban subpopulation, the main focus  of Chinese migration studies (e.g., \cites{Zhao1999, LIU2008, Giles2018, Minale2018, Bairoliya2021}). By combining the 1990, 2000, and 2010 Chinese Censuses with the 1995, 2005, and 2015 1\% population sample surveys, \textcite{Su2018} show that intra-provincial rural migration flows have outnumbered inter-provincial flows since 2011 and empirically identify that the likelihood of moving within the province of origin increases for rural migrants who are older, more educated, female, single, or from poorer areas.\par

\begin{figure}[!htb]	
\centering
\captionsetup{justification=centering}	
\includegraphics[scale=0.5]{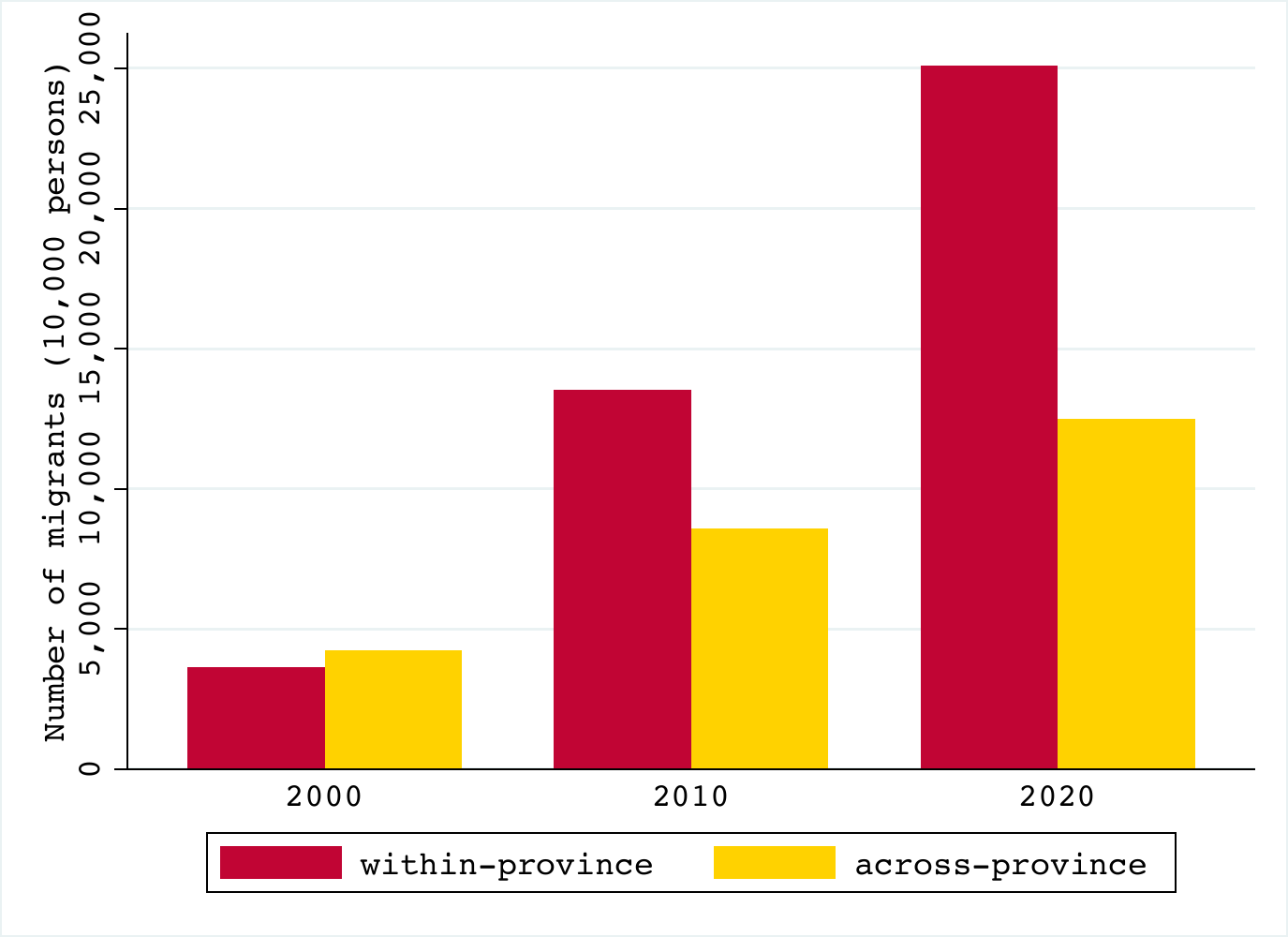}
\caption{\fontsize{10}{12}\selectfont{The number of intra-provincial migrants vs. inter-provincial migrants. \\Source: The 2000, 2010 and 2020 Chinese Censuses.}}
\end{figure}

\hspace{1.4em}Determinants of migration are often conceptualised as push-and-pull factors associated with the sending and receiving locations, respectively. Economic reasons, such as income gaps, job opportunities, and land tenure insecurity, are often the most decisive factors in a substantial body of literature (e.g., \cites{Seeborg2000, Zhu2002, Liu2014, Liu2020, Mullan2011}). However, even though migrants usually move for monetary reasons, contributing factors vary between groups and over time. Demographically, researchers continually find that age, gender, marital status, education level, and family size can explain the probability of migration (e.g., \cites{He2003, Liang2004, Mu2021, Ding2021, Zhao2019}). Spatial and social factors, such as geographic distance, urban amenities, public policies, and the Hukou registration barrier, also play a vital role in propelling or discouraging migration (e.g., \cites{Fan2005, Xing2017, Chau2014, Liu2017, Zhang2020}). The academic debate on the weight of economic opportunities vs. amenities in migration has accordingly received more attention (e.g., \cites{Mueser1995, Liu2014a, Wang2020c}). Additionally, there is an increasing emphasis in the migration literature on environmental degradation, such as climate change and air pollution, caused mainly by industrialisation and over-population (e.g., \cites{ Zhou2011, Liu2020a, Lai2021}).\par

\hspace{1.4em}In addition to these determinants, the impacts of migration on migrants' families, such as consumption patterns and left-behind children (e.g., \cites{Meng2020a, Li2021}), impacts on areas of destination, such as skill aggregation, wage premiums, and occupational upgrading (e.g., \cites{Chung2020, Zhao2020, Combes2020}), and impacts on other potential migrants, such as networks and co-location (e.g., \cites{Foltz2020, Fu2012}), among others, have been widely discussed.\par

\subsection{Expectations in Migration Decision-Making}
\hspace{1.4em}Individual expectations about future outcomes are demonstrated as a compelling driving force behind migration decisions. For instance, \textcite{DEJONG2000} found that income expectancies of remaining in home communities vs. living in alternative locations, in addition to residential satisfaction, were key determinants of migration intentions for both men and women in Thailand. \textcite{Baldwin2001} extended the core-periphery model, illustrating that the prevailing assumption of myopic migrants holds when migration costs are high, while forward-looking expectations arise in scenarios where migration costs are relatively low. More recently, \textcite{Baumann2015} developed a Harris-Todaro model and used US state-level data to show that unemployment rates \textit{per se} did not affect migration, but rather that changes in residents' expectations of unemployment across regions induced migration.\footnote{The Harris-Todaro model stems from \textcite{Harris1970} where migration is driven by urban-rural differences in expected earnings, with the urban employment rate acting as an equilibrating force.} Likewise, \textcite{Shrestha2020} conducted a randomised field experiment to demonstrate that individual expectations changed by gaining information on earnings and mortality rates abroad influenced actual migration decisions of potential migrants in Nepal, particularly of inexperienced migrants.\par

\hspace{1.4em}Other studies of note are of two types. The first is grounded in discrete choice models initiated by \textcite{McFadden1974}. Here, a canonical RUM model includes a deterministic and a stochastic component of utility and a time-specific migration cost. The distributional assumptions on the stochastic term determine the expected probability of selecting a destination. The deterministic term is typically modelled as a function of state variables, such as income, population, or temperature.\footnote{In other words, conventionally, the deterministic component is measured through instantaneous, absolute outcomes.} \textcite{Bertoli2016} expanded the RUM model to allow migrants to make sequential decisions, such as return to their  origin cities or move to alternative locations after migration. They assumed that migrants who are neither myopic nor living in a frictionless world where migration costs are zero respond to the future attractiveness and accessibility of all locations in the choice set. By using two proxies for expectations, they found that economic prospects at origin significantly influenced the scale of migration to Germany. \textcite{Beine2019} also modified the RUM model but contended that agents' expectations of future outcomes were formed by the current level of economic activity and employment rates both within the country of origin and in a number of potential destination countries. Two proxies for signalling expectations about future employment probabilities at destination are found to have been influential in bilateral migration flows.\par

\hspace{1.4em}Another study type considers behavioural theories. \textcite{Czaika2015} developed the migration prospect theory grounded on \posscite{Kahneman1979} seminal work on prospect theory. Prospect theory suggests that the utility of an agent does not depend on an absolute outcome but on gains or losses gradually perceived relative to the reference point, or, in other words, on deviations from the status quo. By applying fundamental principles of prospect theory, i.e., loss aversion, reference-dependence, risk preferences, and diminishing sensitivity, to the migration decision-making model, he demonstrated that migration flows responded more strongly to negative economic and unemployment prospects in home countries than to equal-sized positive prospects in Germany. \textcite{Clark2017} have also gained insights from prospect theory, emphasising the endowment effect whereby people place a higher use value on the object they own than its market value to explain residential moving or staying. Their findings showed that, for internal migration in Australia, the probability of staying increases with stronger risk aversion, is higher for owners than for renters and is higher still with longer duration at the current address.\footnote{Housing tenure, the duration at the current address, and neighbourhood socioeconomic status are the variables used to construct endowment effects.} The work of \textcite{Yonemoto2021} also touches upon prospect theory. He introduced the reference-dependence to the Harris-Todaro model, showing that the reference-dependency can mitigate the over-population in developed regions if people are accurately aware of their post-migration situations and vice versa.\par


\section{Theoretical and Empirical Framework}

\subsection{A Reference-Dependent Migration Model}

\hspace{1.4em}When individuals are visionary, the location-specific utility that reflects the sequential nature of the decision-making problem defined by \textcite{Bertoli2016} based on dynamic discrete choice models \autocites{Artuc2010, Arcidiacono2011, Kennan2011} is as follows:
\begin{gather}
U_{ijk,t}=w_{kt}+\beta A_{kt+1}(I)-c_{jk,t}+\epsilon_{ijk,t}
\end{gather}
where $U_{ijk,t}$ is the utility of an individual $i$ who moves from city $j$ to city $k$ at time $t$. $w_{kt}$ is the deterministic instantaneous component of utility gained by moving from city $j$ to city $k$ at a time $t$. $c_{jk,t}$ describes the cost of moving from $j$ to $k$ at $t$. $\epsilon_{ijk,t}$ is an individual stochastic and time-specific serially-uncorrelated component of utility. Both $w_{kt}$ and $c_{jk,t}$ are known to the individual. The expected utility gained by moving from city $j$ to city $k$ at time $t$ and optimally choosing the preferred location from $t+1$ onward is $A_{kt+1}(I)$. We add $I$ here to denote that the location preference hinges on the industry to which job categories that individual $i$ searches for belong. Jobs sought at destination are not necessarily identical to individual $i$'s previous work at origin. $\beta \in [0, 1)$ is the time discount factor of the expected utility. $\beta=0$ represents the fact that potential migrants do not attach importance to future outcomes and as a result, make myopic decisions. It is also assumed that individual $i$ chooses the preferred location after being aware of the stochastic component of utility at time $t$ for all cities.\par

\hspace{1.4em}\textcite{Beine2019} express the same idea that future outcomes matter, but assume that individual expectations are formed by extracting information from current economic conditions. Their model can be written as:
\begin{equation}
U_{ijk,t}=w_{kt}+\ln (E[A_{kt}(I)])-c_{jk,t}+\epsilon_{ijk,t}
\end{equation}
\hspace{1.4em}Although the model is constructed differently, $E[A_{kt}(I)]$ is basically the same to individual $i$ as the discounted value $\beta A_{kt+1}(I)$ at time $t$, but will be different at some time $t+g\geq t+1$ if individual $i$ moves away from $k$, because the expected instantaneous utility implies a permanent stay or move \autocite{Kennan2011}. Furthermore, Equation (2) takes the log of the expected utility to express the non-linearity and constant relative risk aversion discussed in \textcite{Anderson2011}.\par

\hspace{1.4em}\textcite{Czaika2015} also takes into account the non-linearity of utility and risk attitudes via replacing the utility function over absolute outcomes with the value function over gains and losses relative to a reference point, as does prospect theory. The reference-dependent migration value function derived from \textcite{Koszegi2009} can be written as:
\begin{gather}
V_{it}=V_{it}^{k} - V_{it}^{j} = M(\tilde{y}^k_{it}-\tilde{y}^j_{it})+N(y^k_{it+1},y^j_{it+1}|y^k_{it},y^j_{it}) \\
where \ \ N(\cdot)=(y^k_{it+1}-y^k_{it}-y^j_{it+1}+y^j_{it})^\alpha =(\Delta_{it+1}y^k-\Delta_{it+1}y^j)^\alpha
\end{gather}
where $V_{it}$ is the value of migration for individual i who moves from city $j$ to city $k$ at time $t$. $M(\cdot)$ is the regular component drawn from absolute outcomes in the origin city $\tilde{y}^j_{it}$ and the destination city $\tilde{y}^k_{it}$. In contrast, $N(\cdot)$ is the reference-dependent utility where present economic situations in the origin and destination city, i.e., $y^k_{it}$ and $y^j_{it}$, respectively act as a reference point in adjusting present expectations about the future, i.e., $y^k_{it+1}$ and $y^j_{it+1}$. The superscript α emphasises the non-linearity.\footnote{In this paper, we are only interested in the reference dependence, so for the description of other features, such as $N(\cdot)$ is concave (risk-averse) for expected gains when $N''(\cdot)\leq 0$ for $x>0$ and convex (risk-friendly) for expected losses when $N''(\cdot)>0$ for $x<0$, please see \textcite{Czaika2015}.}\par

\hspace{1.4em}Here, we fine-tune the expected continuation payoff $A_{kt+1}(I)$ in Equation (1) to be reference-dependent, as follows:
\begin{gather}
A_{kt+1}(I)=y^k_{it+1}-y^k_{it}=\Delta_{it+1}y^k (I); A_{jt+1}(I)=y^j_{it+1}-y^j_{it}=\Delta_{it+1}y^j (I)
\end{gather}
\hspace{1.4em}In a one-time migration scenario, we can rewrite Equation (5) as:
\begin{gather}
A_{kt+g}(I)=\frac{1}{\beta^g}E[A_{kt}(I)]=\frac{1}{\beta^g}E[A_{kt}(I)|r^g]\cdot E[r^g]=\frac{r^g}{\beta^g}\cdot \Delta_{it}y^k (I)
\end{gather}
where $r\in (0, 1)$ denotes the uncertainty between present trends and future realisations. $t+g\geq t+1$ is any future point in time from time $t+1$ onward. $r^g \Delta_{it}y^k (I)$ indicates that the further the future, the greater the uncertainty. Similarly, $\beta^g A_{kt+g}(I)$ can be understood as the further the future, the less influential will it be to the present utility.\par

\hspace{1.4em}Nevertheless, the continuation payoff $\Delta_{it+1}y^k (I)$ derived from \textcite{Bertoli2016} entails a more intriguing intuition than the one-time approach that individuals can update their reference points after moving to $k$ at time $t$ and choose any alternative location $q$ among the choice set $D$ at time $t+1$ in terms of their new reference points. If we assume that the stochastic component of utility follows an independent and identically distributed (i.i.d) Extreme Value Type-1 distribution \autocite{McFadden1974} with zero mean where $\tau$ is the Euler constant, the recursive form of the expected utility conditional on residing in $k$ at time $t+1$, in terms of \textcites{Small1981, Kennan2011}, can be expressed as:\footnote{For the reference-independent version, please see \textcite{Bertoli2016}.}
\begin{equation}
\Delta_{it+1}y^k (I)=\tau+\ln\left(\sum\limits_{q\in D}e^{w_{qt+1}-c_{kq,t+1}+\beta\Delta_{it+2}y^q (I)}\right)
\end{equation}
\hspace{1.4em}Then, Equation (1) can be correspondingly rewritten as:
\begin{equation}
U_{ijk,t}=w_{kt}+\beta\left(\tau+\ln\left(\sum\limits_{q\in D}e^{w_{qt+1}-c_{kq,t+1}+\beta\Delta_{it+2}y^q (I)}\right)\right)-c_{jk,t}+\epsilon_{ijk,t}
\end{equation}

\hspace{1.4em}As \textcite{McFadden1974} shows, the probability of migrating from city j to city k can be estimated as:
\begin{equation*}
P_{ijk,t}=Pr\{U_{ijk,t}=\underset{q\in D}{\mathrm{max}} \ U_{ijq,t}\}=\frac{e^{U_{ijk,t}}}{\sum\limits_{q\in D}e^{U_{ijq,t}}}
\end{equation*}
\begin{equation}
In\left(\frac{P_{ijk,t}}{P_{ijj,t}}\right)=w_{kt}-w_{jt}-c_{jk}+\beta\cdot [\Delta_{it+1}y^k (I)-\Delta_{it+1}y^j (I)]
\end{equation}
where $w_{jt}$ is the utility for individual $i$ choosing to remain in city $j$ at time $t$. The probability of $k$ being chosen over $j$ among the choice set $D$ is equivalent to the probability of a binary choice of $k$ over $j$ if we assume the denominator is positive for all possible alternative choices.\footnote{This assumption causes very little loss of generality. Please see \textcite{McFadden1974} for a detailed discussion.}\par

\hspace{1.4em}So far, we do not account for the non-linearity of utility as seen in Equations (2) and (4). It is very plausible that values attached to the migration project are not marginally linear but subject to the scale of change in economic situations, particularly, of the reference point. We can extend Equation (5) by taking the logarithm of $\Delta_{it+1}y^j (I)$ and $\Delta_{it+1}y^k (I)$. Thus, Equation (9) can be rewritten as:
\begin{gather}
In\left(\frac{P_{ijk,t}}{P_{ijj,t}}\right)=w_{kt}-w_{jt}-c_{jk}+\beta\cdot \ln \frac{\Delta_{it+1}y^k (I)}{\Delta_{it+1}y^j (I)}
\end{gather}

\subsection{Econometric Modeling and Techniques}
\hspace{1.4em}Our core predictor, as the empirical counterpart of $\Delta_{it+1}y^j (I)$ and $\Delta_{it+1}y^k (I)$, is a proxy for job prospects in the city of origin and city of potential destination, respectively. We use the term `trending' to highlight the fact that it captures upward and downward trends at the industry level. Macro surroundings often silently yet profoundly influence individual perceptions and accordingly, trending signals derived here mirror the role of contextual evolution in forming expectations of all relevant individuals.
\begin{gather}
Job\_Trending_{ij,t}=\frac{E_{ij,t}-E_{ij,t-1}}{E_{ij,t-1}}-\frac{E_{ij,t-1}-E_{ij,t-2}}{E_{ij,t-2}}=GR_{ij,t}-GR_{ij,t-1}=\Delta_{t} GR_{ij} \\ 
Job\_Trending_{ik,t}=\frac{E_{ik,t}-E_{ik,t-1}}{E_{ik,t-1}}-\frac{E_{ik,t-1}-E_{ik,t-2}}{E_{ik,t-2}}=GR_{ik,t}-GR_{ik,t-1}=\Delta_{t} GR_{ik}
\end{gather}
where the quantity of employment at time $t$ in the sector of job categories that individual $i$ looks for is $E_{ij,t}$ for the origin city $j$ and $E_{ik,t}$ for the destination city $k$. In short, the trending indicator is the annual change in growth rates of industrial employment, i.e., $\Delta_{t} GR_{ij}$ and $\Delta_{t} GR_{ik}$. The beauty of this design is that positive growth does not necessarily lead to better job prospects and vice versa. If we consider that $GR_{ij,t}=1.5\%$ and $GR_{ij,t-1}=1.8\%$, despite both being positive, the outcome is $-0.3\%$ signalling a slowdown. Likewise, for negative growth over two years, such as $GR_{ij,t}=-0.5\%$ and $GR_{ij,t-1}=-0.7\%$, $\Delta_{t} GR_{ij}=0.2\%$ is still a positive value, because the scale of the present decline is narrower, implying that job prospects stand a chance of getting better. In other words, individuals gain utility from staying or moving, not because of the growth \textit{per se}, but due to whether progress is faster (better) or slower (worse) relative to the previous year at origin vs. destination.\par

\hspace{1.4em}The implication for this construction inspired by the migration prospect theory \autocite{Czaika2015} can be linked to \posscite{Baumann2015} findings that unemployment affects migration only if it alters expectations, and the central concept of \textcite{Clark2017}. In our context, it could be understood that $GR_{ij,t}=1.5\%$ does not trigger migration since no expected gain or loss emerges if $GR_{ij,t-1}$ also $=1.5\%$. Moreover, \textcite{Clark2017} elaborate what prospect theory could offer for understanding why the majority of people prefer staying, because, in terms of loss aversion, people do not necessarily choose the highest expected utility gained via migration if they are more concerned about losing what they have. This offers insights into understanding our indicator from another angle. If we consider an industry with consistent growth at both origin and destination, to compete against the status quo bias ingrained in resident workers, deviations from two reference points are critical to add weight to moving and, consequently, to impel decision-making.\footnote{We can elaborate `deviations from reference points' with two questions. First, are future situations at origin better than the present? Negative changes triggering loss aversion could strongly encourage emigration. Second, are changes at destination better than at origin? A larger scale of expansion at destination is required particularly when positive changes are also observed in origin cities.}\par

\hspace{1.4em}Thus, we can formulate the distance of expected utility between the destination and origin city:
\begin{gather}
Distance(Job\_Trending_{ijk,t})=\Delta_{t} GR_{ik}-\Delta_{t} GR_{ij}
\end{gather}
\hspace{1.4em}The distance variable corresponding to Equation (10) can be calculated as:
\begin{gather}
Distance^T(Job\_Trending_{ijk,t})=\ln(\Delta_{t} GR_{ik})-\ln(\Delta_{t} GR_{ij})=\ln(\frac{\Delta_{t} GR_{ik}}{\Delta_{t} GR_{ij}})
\end{gather}
where $Distance^T$ is the transformed version of Equation (13). The assumption is that holding the difference in job prospects constant, potential migrants are more reluctant to move when their expectations of job prospects in their origin cities are good, and the better the prospects, the less responsive potential migrants become  (e.g., given $\Delta_{t} GR_{ik}-\Delta_{t} GR_{ij}=1$, $\ln(2)-\ln(1)=0.69$ vs. $\ln(10)-\ln(9)=0.11$). Conversely, the worse the local prospects, the more susceptible potential migrants become.\footnote{Because the trending indicator contains negative values, we applied the so-called `started logarithm' \autocite{Tukey1977}, i.e., $\ln(y+c)$ where $c>0$ is set such that $y+c>0$ for all $y$, to transform the variable. We estimated both versions and report results of the untransformed distance predictor in the Appendix.}\par

\hspace{1.4em}As binary outcomes cannot be log-transformed, our fixed-effects specifications act as a linear probability model approximation to Equation (10), which can be accurate in magnitude and sign for small parameter values in practical matters \autocite{McFadden1974}. The baseline to be estimated is as follows:
\begin{equation}
M_{ijk,t}=\alpha+\beta_{1}X_{it}+\beta_{2}Distance^T(Job\_Trending_{ijk,t})+\beta_{3}Distance(Z_{ijk,t-1})+\gamma_{t}+\epsilon_{ijk,t}
\end{equation}
where the binary dependent variable (DV) $M_{ijk,t}$ equals 1 if individual $i$ chooses to move from city $j$ to city $k$ at time $t$, and 0 otherwise. $X_{it}$ is a row vector of individual and household characteristics. $Distance(Z_{ijk,t-1})=Z_{ik,t-1}-Z_{ij,t-1}$ are city-level control variables. Because these covariates are state variables measured at the end of each year, we lag them by one period. $\gamma_{t}$ are time-fixed effects and $\epsilon_{ijk,t}$ is an idiosyncratic error term.\par

\hspace{1.4em}In this paper, the city covariates refer to income level, usually measured through GDP per capita \autocite{Beine2016a}, population density in relation to traffic congestion and informal settlement and slums \autocites{Neirotti2014, Tan2016}, and urban amenities associated with two major aspects: the provision of healthcare (the number of hospital beds per person) and higher education resources (the number of higher educational institutions per 10,000 persons) \autocite{Czaika2017}. We also include the share of the tertiary sector and the number of enterprises above the designated size per 10,000 persons to account for gaps in regional commercialisation and business density, respectively.\footnote{The term `enterprises above the designated size' was first used in China in 1996 and defined by the National Bureau of Statistics. Before 2011, firms with an annual output of 5 million Yuan or more were counted as enterprises above the designated size. The benchmark has raised up to 20 million Yuan since 2011.} Additionally, in some specifications, we add the China Hukou Registration Index (CHRI) developed by \textcite{Zhang2019a} for 120 Chinese cities to control for Hukou entry policies and migration costs.\footnote{The CHRI divided the local Hukou registration policies into four overarching aspects -- talent recruitment, general employment, investment and taxation as well as home purchase and then constructed the index for two stages (2000--2013 and 2014--2017).} The individual and household covariates are: gender, marital status, Hukou type (rural or urban), self-evaluated health status, household income, age, years of schooling, and family migration network. Research suggests that the presence of networks lowers migration costs, increases the probability of migration and is more important for the mobility of low-educated migrants than for the higher-educated (e.g., \cites{Orrenius2005, Mckenzie2007, Beine2011}). We construct it as a dummy, indicating if there are any pioneer migrants within families prior to the move of each individual.\par

\hspace{1.4em}We then add three fixed effects to Equation (15) to alleviate endogeneity issues, as follows:
\begin{align}
\begin{split}
M_{ijk,t}=\alpha+\beta_{1}X_{it}+\beta_{2}Distance^T(Job\_Trending_{ijk,t})+\beta_{3}Distance(Z_{ijk,t-1}) \\+\gamma_{t}+\gamma_{j}+\gamma_{k}+\gamma_{s}+\epsilon_{ijk,t}
\end{split}
\end{align}
where $\gamma_{j}$, $\gamma_{k}$ and $\gamma_{s}$ are origin, destination and industrial sector fixed effects. Adding them helps reduce unobservable time-invariant or very slowly varying push and pull factors of each origin, destination, and industrial sector, such as the Hukou registration policies in origin or destination cities, industry-specific policies, or demographic characteristics of the population in each city or industry. This is the traditional strategy to control for multilateral resistance to migration in cross-sectional studies \autocite{Mayda2010}. In such cases, multilateral resistance to migration occurs when a destination city has different levels of attractiveness to people from the same place of origin due to gender, age, educational level, etc. This heterogeneity implies that origin-specific patterns of correlation across all potential locations exist in the stochastic component of utility.\par 

\hspace{1.4em}However, the absence of future attractiveness of alternative locations which has an impact on choices of moving between $j$ and $k$ can still bias the estimation \autocite{Bertoli2013}. More specifically, assume that migration between Suzhou and Shanghai increases because the movers have worse expectations about their job prospects in Suzhou, and these worsening prospects are correlated with their worsening job prospects in Nanjing. Then, if the influence of alternative locations is not considered, the increase in the bilateral migration to Shanghai would be wholly attributed to worsening job prospects in Suzhou, resulting in overestimation. In other words, failure to account for multilateral resistance to migration might entail a violation of the IIA hypothesis underlying the discrete choice model discussed above.\par

\hspace{1.4em}Ideally, the solution to multilateral resistance to migration is  \posscite{Pesaran2006} common correlated effects (CCE) estimator if the cross-sectional and longitudinal dimensions of the panel are large enough. Yet our dataset does not meet this computing demand, so we follow the less data-demanding approach used in a wide range of applied migration literature (e.g., \cites{Ortega2013, Beine2015, Royuela2018, Maza2019}).\par

\hspace{1.4em}We change Equation (16), first by replacing monadic fixed effects of origin and destination with dyadic fixed effects of origin-destination, as follows:
\begin{align}
\begin{split}
M_{ijk,t}=\alpha+\beta_{1}X_{it}+\beta_{2}Distance^T(Job\_Trending_{ijk,t})+\beta_{3}Distance(Z_{ijk,t-1})\\+\gamma_{t}+\gamma_{jk}+\gamma_{s}+\epsilon_{ijk,t}
\end{split}
\end{align}
\hspace{1.4em}This specification is similar to the above but allows us to control for deterministic effects for each pair of cities. $\gamma_{jk}$ can capture specific bilateral migration relationships between $j$ and $k$, such as geographic distance, migration costs, or historic migration networks. Other variables remain the same, as above.\par

\hspace{1.4em}One of the most common approaches to dealing with multilateral resistance to migration is to apply dyadic fixed effects of origin-time, as follows:
\begin{align}
\begin{split}
M_{ijk,t}=\alpha+\beta_{1}X_{it}+\beta_{2}Distance^T(Job\_Trending_{ijk,t})+\beta_{3}Distance(Z_{ijk,t-1})\\+\gamma_{jt}+\gamma_{k}+\gamma_{s}+\epsilon_{ijk,t}
\end{split}
\end{align}
where $\gamma_{jt}$ is a vector of origin dummies for each year. All other variables remain the same as in the main specification. In terms of \textcites{Royuela2018, Ortega2013}, this method enables us to control for all the push determinants of migration decisions and particularly, multilateral resistance derived from heterogeneity in migration preferences that are constant across destination cities and that vary only by year and city of origin. Previous literature where DV was measured though migration flows also used $\gamma_{jt}$ to account for the denominator $P_{ijj,t}$ \autocite{Beine2016a}.\par

\hspace{1.4em}Another common approach is to use dyadic fixed effects of destination-time in the place of origin-time, as follows:
\begin{align}
\begin{split}
M_{ijk,t}=\alpha+\beta_{1}X_{it}+\beta_{2}Distance^T(Job\_Trending_{ijk,t})+\beta_{3}Distance(Z_{ijk,t-1})\\+\gamma_{kt}+\gamma_{j}+\gamma_{s}+\epsilon_{ijk,t}
\end{split}
\end{align}
where $\gamma_{kt}$ is a vector of destination dummies for each year. All other variables remain the same as in the main specification. As explained in \textcite{Beine2015, Royuela2018}, this strategy allows us to control for all the pull determinants of migration decisions and dynamic resistance derived from heterogeneity in the future attractiveness that are constant across origin cities and that vary only by year and city of destination.\par

\hspace{1.4em}Based on Equations (18)--(19), we replace the monadic fixed effects of industrial sector with dyadic fixed effects of industry-time because our core predictor rests on sectors:
\begin{align}
\begin{split}
M_{ijk,t}=\alpha+\beta_{1}X_{it}+\beta_{2}Distance^T(Job\_Trending_{ijk,t})+\beta_{3}Distance(Z_{ijk,t-1})\\+\gamma_{jt}+\gamma_{k}+\gamma_{st}+\epsilon_{ijk,t} \\ 
M_{ijk,t}=\alpha+\beta_{1}X_{it}+\beta_{2}Distance^T(Job\_Trending_{ijk,t})+\beta_{3}Distance(Z_{ijk,t-1})\\+\gamma_{kt}+\gamma_{j}+\gamma_{st}+\epsilon_{ijk,t}
\end{split}
\end{align}
where $\gamma_{st}$ is a vector of industry dummies for each year, capturing heterogeneity in migration preferences that vary by year and industrial sector, such as emerging jobs. All other variables remain the same as in the main specification.

\hspace{1.4em}As we have seen, the less data-demanding solution to multilateral resistance to migration relies on utilising various structures of fixed effects. Nevertheless, a fixed effects estimator can be significantly biased in non-linear models \autocite{Neyman1948}. \textcite{Beck2018} demonstrates that the larger the number of fixed effects, the stronger bias imposed on fixed effects logit models. In contrast, the critique of using a linear probability model (LPM) is usually twofold: a) predicted probabilities might be negative or above 1, such as 1.2 or -0.4, that are unrealistic, and b) the dichotomous DV renders heteroskedasticity which violates one of the OLS assumptions that all disturbances have the same variance. Yet the heteroskedasticity can be addressed with heteroskedasticity-consistent robust standard errors, while the first concern is most often why the LPM is not preferred. \textcite{Horrace2006} shows that the bias and inconsistency of LPM increase with a greater proportion of predicted probabilities falling outside the unit interval. In other words, if all predicted probabilities fall between 0 and 1, the linear probability estimator can be unbiased and consistent (e.g., \cite{Friedman2013}).\par

\hspace{1.4em}In our sample, 8.3\% of observations present a negative predicted migratory probability drawn from Equation (16) of which only 13\% are migrants. We thus report estimates of the sub-sample where all predicted probabilities are within the unity as a supplement. Further, we consider a multilevel logistic regression as part of the robustness check. It enables us to take into account context-specific influences on potential migrants in accordance with the theoretical perspective that people from the same place of origin tend to behave more similarly than do individuals from other places due to a variety of spatial and socio-cultural proximity. \textcite{DAgostino2019} shows that the multilevel method is useful to control for unmeasured area-specific effects interrelated with migration propensities.\par

\hspace{1.4em}Let $\pi_{it}=Pr(M_{ijk,t}=1)$, our two-level logistic random intercept model is formulated as follows and purely an empirical counterpart of Equation (10):
\begin{align}
\log (\frac{\pi_{it}}{1-\pi_{it}})=\alpha+\beta_{1}X_{it}+\beta_{2}Distance^T(Job\_Trending_{ijk,t})+\beta_{3}Distance(Z_{ijk,t-1})+\gamma_{t}+\mu_{C}
\end{align}
where $\mu_{C}$ is assumed to be i.i.d normally distributed with a zero mean and level-2 variance $\sigma_{C}$, accounting for the effects of being in city group $C$ on the log-odds that $M_{ijk,t}=1$. $\beta_{1}$ is still level-1 unknown parameters, and $\beta_{2}$ and $\beta_{3}$ are level-2 parameters to be estimated. The index $C$ can be either an origin city or a destination city. Later, we estimate both.\par

\hspace{1.4em}We can extend the two-level model to a three-level model if we postulate that within the same city group, the type of contextual information that people access and the way they are impacted differ across education levels. In this case, city effects $\mu_{C}$ become level-3 and a new level-2 random intercept, i.e., $\mu_{Cv}$, is added to Equation (21). The index $v$ denotes the year of schooling.\par

\hspace{1.4em}In addition, \textcite{Wintoki2012} show that the generalised method of moments (GMM) estimation yields better results than a fixed effects estimator for at least two potential issues of endogeneity: unobserved heterogeneity and simultaneity. On the basis of \textcite{Hansen1982}, the two-step system GMM is developed by \textcite{Arellano1995} and \textcite{Blundell1998} where lagged first differences are utilised as instruments in the level equation at the cost of an additional assumption that first differences of instrument variables are uncorrelated with unobserved unit-specific heterogeneity. We, thus, further continue our analysis using a system GMM estimator.\footnote{It is unreasonable to assume that the previous migration decision affects the present migration decision. We therefore do not consider adopting a lagged DV.} \textcite{Bellemare2017} compares the bias and consistency of three estimators with a static panel model and find that the GMM technique always outperforms OLS regression.\footnote{\textcite{Rahko2021} recently shows an underestimate of the OLS estimator compared to the static system GMM results.} The model is almost the same as Equation (15) but with an inclusion of industry fixed effects.\par

\subsection{Data Description}

\subsubsection{Data Source}
\hspace{1.4em}The quasi-panel is created by combining a nationally representative cross-sectional micro-data, the 2017 China Household Finance Survey (CHFS), with city-level longitudinal statistics during 1997--2017, retrieved from the China Data Institute. This approach is not uncommon for applied research.\footnote{To give an example, \textcite{Yamada2020} created a quasi-panel to study the impact of Myanmar's township-level conflicts on education, and \textcite{Schmidt-Catran2016} exploited hybrid methods to investigate cross-sectional as well as longitudinal effects of migration on native Germans' support for welfare.} The CHFS is one of the very few nationwide surveys that allow us to identify where migrants are from and where they settle at the city level and the year in which their migration occurs.\footnote{Another option is the CMDS, however, we do not have permission to access it.} After comprehensive data cleaning and compiling, our panel eventually contains 10,254 migrants and 56,173 natives. For details of the data preparation, please see \hyperref[tab9]{Appendix C}.\par

\hspace{1.4em}It is worth noting again that all individual and household variables were retrieved from the 2017 wave. After merging the cross-sectional survey data with the longitudinal statistics, we re-calculated the age and years of schooling of each individual for each year and only retained observations of ages between 16 and 65. As a result, we finally have three time-varying variables at this level, that is: age, years of schooling, and family migration network. As migrants are defined as those who have moved across prefecture boundaries, and to focus on the decision-making problem, our binary DV is only equal to 1 in the year of migration and to 0 at all other points in time. Furthermore, we distinguished individuals who had moved and transferred their Hukou to destination cities from natives and specified them as migrants. As migrants in China are often measured by the separation of the Hukou and resident places, people who transferred the Hukou to their cities of destination are treated as natives in the majority of migration studies \autocite{Zhang2020}. However, ignoring those migrants with Hukou transfer leads to an underestimate of the scale of migration.\par

\subsubsection{Descriptive Statistics}
\hspace{1.4em}In our RUM model, we discuss that the expected utility $A_{kt+1} (I)$ is conditional on the industry in which job categories that individual $i$ seek are mainly based. From the CHFS, we learn each individual's industry group, if he or she has a formal job. Table 1 classifies the industry groups based on the Chinese national standard number `GB/T 4754' and presents corresponding statistics of each group. As shown, at the aggregate level, the primary sector has the lowest proportion of labour, as opposed to the tertiary sector. Among sub-sectors, 21.3\% of individuals in our sample were employed in Manufacturing, followed by Construction (9.7\%), Social Services (9.1\%) and Wholesale and Retail Trade (9\%).\par

\hspace{1.4em}The trending indicator is eventually quantified by the statistics of the three main sectors. Approximately 25,000 natives and 3,000 migrants in our sample do not provide industry groups of their jobs because they are non-employee workers. Thus, total employment statistics are used for this group instead of sectoral statistics. Accordingly, we have four sets of industry-specific dummies applied in our estimation (primary, secondary, tertiary and total). Furthermore, Figure 3 is the geographic distribution of emigrants employed in the secondary sector moving across  province boundaries, from which we see five Chinese provinces suffered the greatest labour outflux: Sichuan, Hubei, Anhui, Hunan, and Guangdong. Similarly, Figure 4 is the plot looking at the tertiary sector. Here, Guangdong still sent out the most migrants followed by Sichuan, Hubei, Anhui, Hunan, and Liaoning. In \hyperref[figb1]{Figure B2}, we also graphically present where migrants are from, defined by the prefecture boundary.\par

\vspace{0.3cm}
\begin{ThreePartTable}
\linespread{1}
\begin{TableNotes}[flushleft]\footnotesize
\item Source: Authors' calculation using CHFS.\par
\end{TableNotes}

\def\sym#1{\ifmmode^{#1}\else\(^{#1}\)\fi}
\begin{spacing}{0.7}
\fontsize{10}{12}\selectfont{
\begin{longtable}{>{\raggedright}p{9cm} c c}
\captionsetup{belowskip=0pt,aboveskip=1pt}
\caption{Industrial Classification \& Statistics\label{tab7}}\\
\toprule\endfirsthead\midrule\endhead\midrule\endfoot\endlastfoot
          {Industry Group} &{Number of Obs.}&{Main Sector}\\
\midrule
Farming, Forestry, Animal Husbandry  & 2695 & Primary Sector \\

Mining and Quarrying & 488 & Secondary Sector \\

Manufacturing & 7765 & Secondary Sector \\

Electric Power Gas and Water Production and Supply & 1184 & Secondary Sector \\

Construction & 3517 & Secondary Sector \\

Wholesale and Retail Trade  & 3294 & Tertiary Sector \\

Transportation Storage Post and Telecommunications  & 2531 & Tertiary Sector \\

Hotel and Catering Services  & 1959 & Tertiary Sector \\

Information Transmission, Software and Information Technology & 870 &  Tertiary Sector \\

Banking and Insurance & 1015 & Tertiary Sector \\

Real Estate & 466 & Tertiary Sector \\

Leasing and Business Services  & 333 & Tertiary Sector \\

Scientific Research, Technical Service and Geologic Prospecting  & 234 & Tertiary Sector \\

Management of Water Conservancy, Environment and Public Facilities & \multirow{2}{*}{455}  & \multirow{2}{*}{Tertiary Sector} \\

Social Services  & 3314 & Tertiary Sector \\

Education & 2112 & Tertiary Sector \\

Health, Social Security and Social Welfare & 1562 & Tertiary Sector \\

Culture, Sports and Entertainment & 613 & Tertiary Sector \\

Public Management and Social Organisation & 1942 &  Tertiary Sector \\
\midrule
\textit{Total} & 36349 &  \\
\bottomrule
\insertTableNotes 
\end{longtable}
}
\end{spacing}
\end{ThreePartTable}

\begin{figure}[!htb]	
\centering
\captionsetup{justification=centering}	
\includegraphics[height=3.5in,width=5.2in]{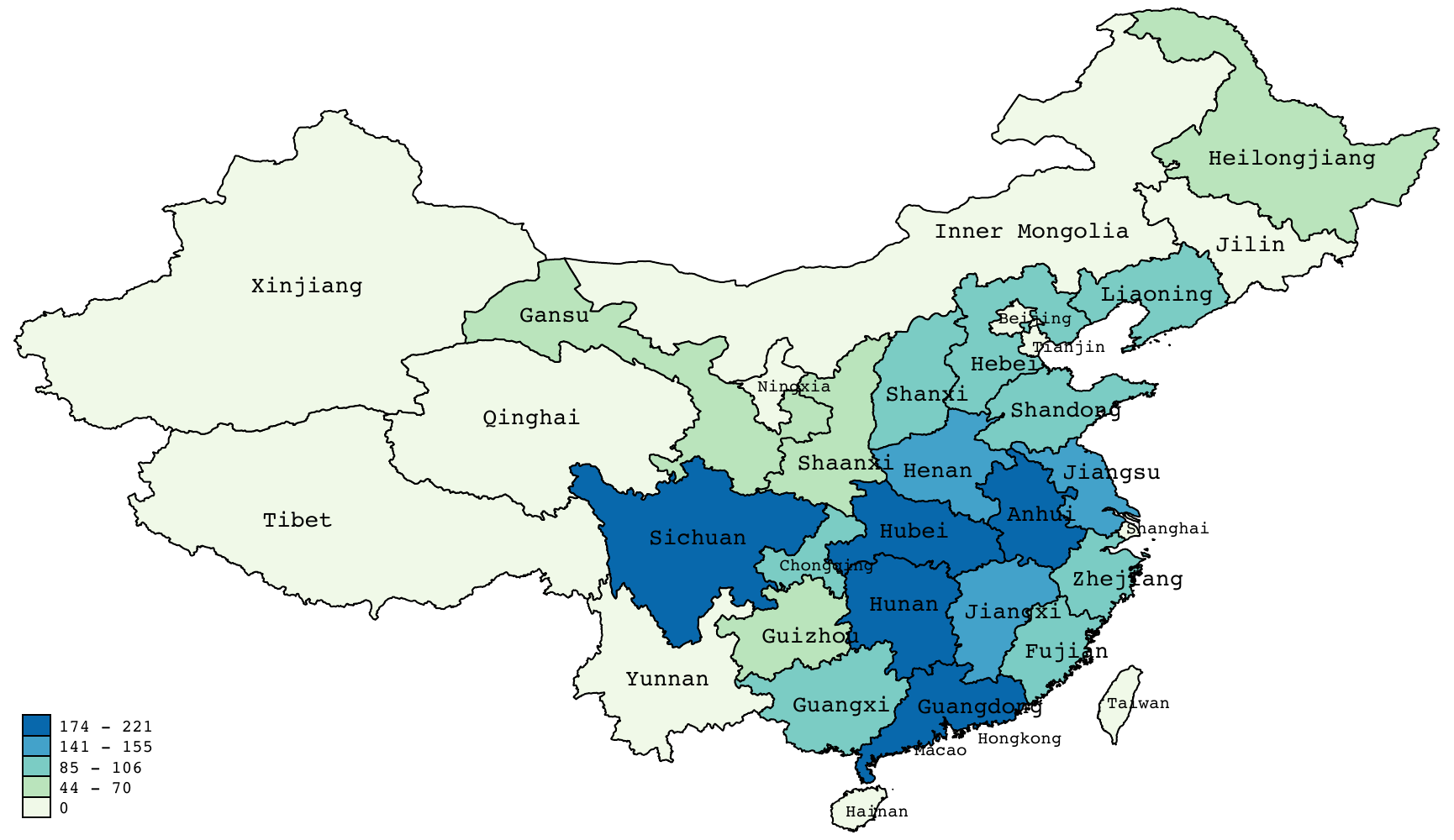}
\caption{\fontsize{10}{12}\selectfont{Province boundary map -- emigration distribution (secondary sector). \\Source: Author's elaboration using CHFS.}\label{fig1}}
\end{figure}

\begin{figure}[!htb]	
\centering
\captionsetup{justification=centering}	
\includegraphics[height=3.5in,width=5.2in]{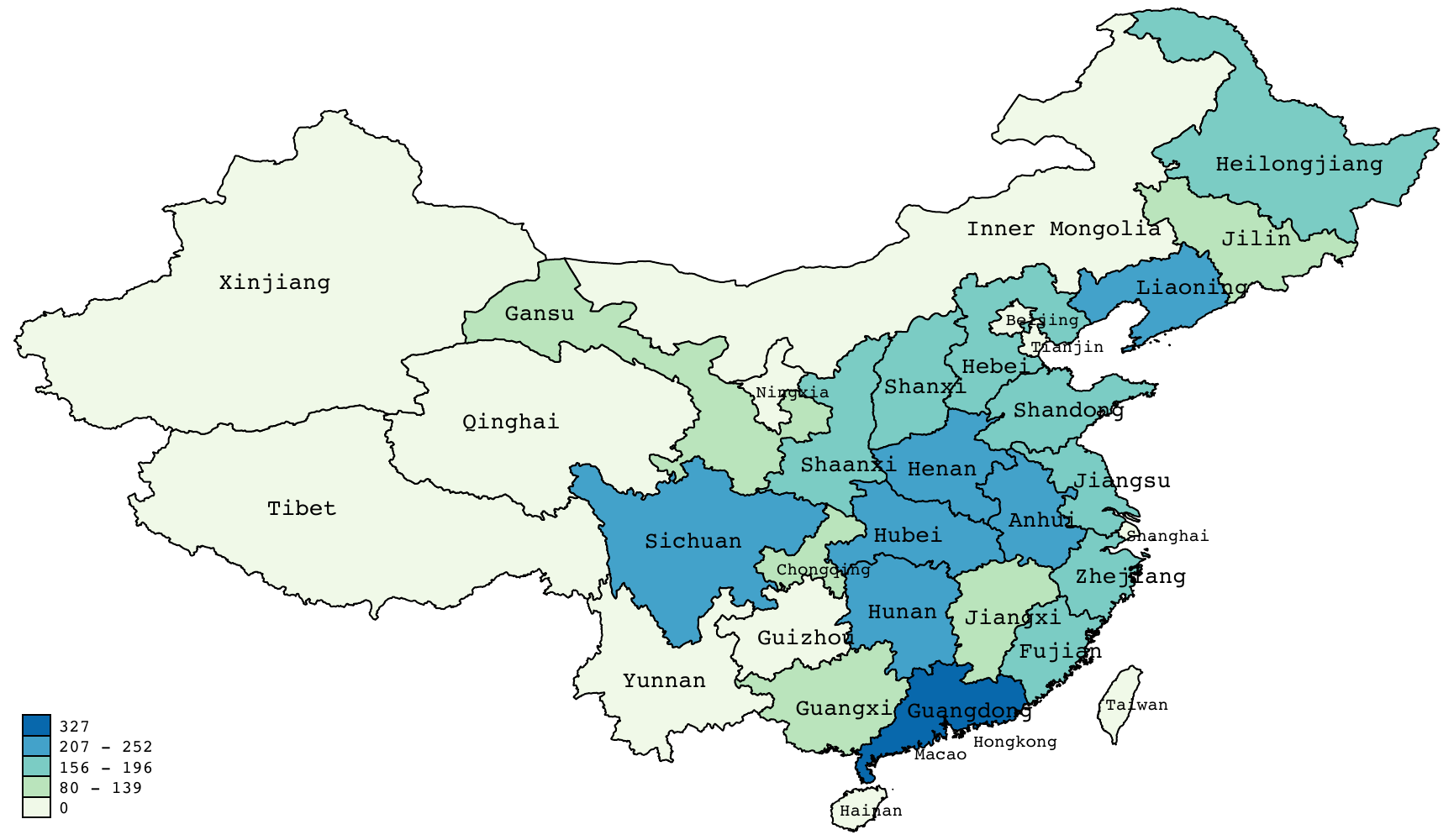}
\caption{\fontsize{10}{12}\selectfont{Province boundary map -- emigration distribution (tertiary sector). \\Source: Author's elaboration using CHFS.}\label{fig2}}
\end{figure}

\FloatBarrier

\hspace{1.4em}Table 2 contains descriptive statistics of the survey data and city statistics, respectively, as well as the DV and the trending indicator ($Distance^T(Job\_Trending_{ijk,t})$) in the final quasi-panel.\footnote{We acknowledge several limitations here. Firstly, as a quasi-panel, most of the individual and household variables are time-invariant. Although the CHFS has four waves, about one-third of migrant samples were new to 2017, and earlier waves still do not have information prior to 2011. Secondly, the migration history we can learn is incomplete. It is possible for migrants to move several times within a year, however, we can only identify their most recent destination. Thirdly, industries to which individuals’ jobs belong depend on their most recent employment. Mismatch between what we learn from the 2017 wave and actual examples that individuals evaluated is possible. Otherwise, examining job prospects at the sub-industry level would be even more interesting.} More specifically, migrants in our sample were on average 11 years younger, had 1.4 more years of schooling, and obtained 6,800 Yuan more in annual wages after-tax than did natives. The gender ratio of females to males among natives was 46:54, but 42:58 among migrants. As to the Hukou type, 44\% of natives, as opposed to 34\% of migrants, had an urban Hukou.\footnote{Prefecture cities in China usually consist of several districts and counties. They have the dual functions of administering both rural and urban areas. An increasing population with the rural Hukou lives in urban areas because of local urbanisation and within-prefecture migration \autocite{Song2014}.} Family sizes did not seem to differ between natives and migrants (1.87 versus 1.89), but fewer migrants were in a marital relationship. Self-evaluated health status shows that migrants overall evaluated their health as slightly better than that of natives.\par

\hspace{1.4em}Moreover, compared to natives, migrants were mostly from economically underdeveloped cities where the GDP per capita was on average 4,600 Yuan less in 2000 and 12,500 Yuan less in 2010 than for natives' resident cities. Generally, migrants moved to economically developed cities where the GDP per capita was 9,800 Yuan more in 2000 and 50,000 Yuan more in 2010 than their areas of origin. Likewise, destination cities usually had 9\% higher ratios of the tertiary sector to the total than cities of origin and three to four times more enterprises above the designated size per 10,000 persons. These destination cities were also on average 41\% more densely populated than migrants' cities of origin but with better public medical and higher education resources per capita. In addition, natives' resident cities applied more stringent Hukou registration policies: the CHRI was 0.17 higher before 2013 and 0.28 higher since 2014 than migrants' cities of origin. However, the Hukou barrier did not indeed prevent migrants from moving to higher indexed cities.\footnote{Actually, their cities of destination were on average 0.37 higher before 2013 and 0.49 higher since 2014 than their cities of origin.} This is mainly because cities setting higher registration barriers were, in general, more economically developed \autocite{Zhang2019a}. Lastly and importantly, although the percentage change in growth rates of industrial employment in origin cities fared slightly better than destination cities, for example, 4.7\% at origin vs. 3\% at potential destination in 2000 and 1.6\% at origin vs. 0.9\% at potential destination in 2010, the situation completely reversed at the time that migration occurred: 0.8\% at origin vs. 2.2\% at destination, on average.\par

\clearpage
\begin{ThreePartTable}
\linespread{1}
\begin{TableNotes}[flushleft]\footnotesize
\item Notes: Approx. 10,000 observations in the 2017 CHFS survey cannot match with city statistics because their resident cities are not included in the statistics. 1,466 households answered with zero or even negative annual income due to debts and losses on investment, among others. The variable `distance\_JobTrend' is the transformed, as defined in Equation (14). The statistical currency is Yuan.
\item Source: Authors' elaboration using CHFS, CHRI and China Data Institute (2021).
\end{TableNotes}

\def\sym#1{\ifmmode^{#1}\else\(^{#1}\)\fi}
\begin{spacing}{0.7}
\fontsize{9}{12}\selectfont{
\begin{longtable}{l>\raggedright p{4cm} c c c c c}
\captionsetup{belowskip=0pt,aboveskip=1pt}
\caption{Descriptive Statistics}\\
\toprule\endfirsthead\midrule\endhead\midrule\endfoot\endlastfoot
          {Variable}          &{Description}&{Obs.}&{Mean}&{Std. Dev.}&{Min}&{Max}\\
\midrule
\textit{CHFS 2017} \\
\midrule
migrant & If native=0; migrant=1	& 75,099	& 0.17	& 0.38	& 0	& 1 \\
\addlinespace
gender& Female=0; Male=1	& 75,099	& 0.55	& 0.50 & 0	& 1 \\ 
\addlinespace
age &	Age in 2017	& 75,099	& 47.82	 & 17.66	& 16 &	85 \\
\addlinespace
\multirow{2}{*}{education} &	Educational level: no schooling=1, PhD=9	& \multirow{2}{*}{74,894}	& \multirow{2}{*}{3.48}	& \multirow{2}{*}{1.75}	& \multirow{2}{*}{1}	& \multirow{2}{*}{9} \\
\addlinespace
marriage	& Married=1; Otherwise=0	& 72,835	& 0.77	& 0.42	& 0	& 1 \\
\addlinespace
hukou\_type	& Urban Hukou=1; Rural=0	  &74,961	& 0.41	& 0.49& 0	& 1 \\
\addlinespace
\multirow{2}{*}{health\_status} &	Health degree: very good=1, very bad=5	 & \multirow{2}{*}{75,072}	& \multirow{2}{*}{2.52} &	\multirow{2}{*}{1.03}	& \multirow{2}{*}{1}	& \multirow{2}{*}{5} \\
\addlinespace
hh\_income	& Log of household income	 & 75,099	& 13.89	& 0.36	& -9.21 	& 15.61 \\
\addlinespace
\multirow{2}{*}{family\_size} & The number of family members living together & \multirow{2}{*}{12,214} & \multirow{2}{*}{1.88} & \multirow{2}{*}{1.28} & \multirow{2}{*}{1} & \multirow{2}{*}{10} \\
\addlinespace
wage & After-tax wage last year & 24,736 & 37533.76  & 45066.41  & 20 & 3100000 \\
\midrule
\textit{City Statistics} \\
\midrule
all\_GR	& Growth rates in all sectors	 & 5,522	& 0.0208	& 0.4284	&-0.9673	& 18.36 \\
\addlinespace
\multirow{2}{*}{prima\_GR}	& Growth rates  in the primary sector 	& \multirow{2}{*}{5,230}	& \multirow{2}{*}{0.1581}	& \multirow{2}{*}{3.55}	& \multirow{2}{*}{-1}	& \multirow{2}{*}{120.96} \\
\addlinespace
\multirow{2}{*}{second\_GR}	& Growth rates in the secondary sector & \multirow{2}{*}{5,738}	& \multirow{2}{*}{0.0132}	& \multirow{2}{*}{0.4543}	& \multirow{2}{*}{-0.9267}	& \multirow{2}{*}{22.72} \\
\addlinespace
\multirow{2}{*}{tertiary\_GR}	& Growth rates in the tertiary sector	& \multirow{2}{*}{5,738}	& \multirow{2}{*}{0.0117}	& \multirow{2}{*}{0.3121}	& \multirow{2}{*}{-0.9395}	& \multirow{2}{*}{7.78} \\
\addlinespace
ppDen & Log of population density	& 5,667	& 5.85	& 1.01	 & 1.55	& 9.55 \\
\addlinespace
Ingdppc	& Log of GDP per capita	&  5,492	& 9.73	& 0.9582	& 7.01 & 12.58 \\
\addlinespace
\multirow{3}{*}{coop}& The number of enterprises above the designated size per 10,000 persons & \multirow{3}{*}{5,425}	& \multirow{3}{*}{2.53}	& \multirow{3}{*}{3.32}	& \multirow{3}{*}{0.12}	& \multirow{3}{*}{36.29}  \\
\addlinespace
\multirow{2}{*}{medical}	&Number of beds in hospitals per person	&\multirow{2}{*}{5,705}	& \multirow{2}{*}{0.0033}	& \multirow{2}{*}{0.0016}	& \multirow{2}{*}{0}	& \multirow{2}{*}{0.0189} \\
\addlinespace
\multirow{2}{*}{highEdu}	& Number of higher educational institutions per 10,000 persons	&\multirow{2}{*}{5,739}	&\multirow{2}{*}{0.0155}	&\multirow{2}{*}{0.0194}	&\multirow{2}{*}{0}	&\multirow{2}{*}{0.1197} \\ 
\addlinespace
\multirow{2}{*}{tertiaryRatio}	& The ratio of output values of the tertiary sector 	&\multirow{2}{*}{5,780}	&\multirow{2}{*}{0.6529}	&\multirow{2}{*}{0.1553}	&\multirow{2}{*}{0.2029}	&\multirow{2}{*}{0.9897} 
\\
\addlinespace
\multirow{2}{*}{CHRI}	& The Hukou registration  stringency index 	&\multirow{2}{*}{2,167}	&\multirow{2}{*}{0.6614}	&\multirow{2}{*}{0.3370}	&\multirow{2}{*}{0.1331}	&\multirow{2}{*}{2.63} 
\\
\midrule
\textit{Merged} \\
\midrule
\multirow{2}{*}{migrate} & =1 at the time of migration; =0 before and afterwards & \multirow{2}{*}{1,128,624} & \multirow{2}{*}{0.0084} & \multirow{2}{*}{0.0914} & \multirow{2}{*}{0} & \multirow{2}{*}{1} \\
\addlinespace
\multirow{2}{*}{distance\_JobTrend} 	& The ratio of job prospects at destination to origin 	 & \multirow{2}{*}{969,998}	&\multirow{2}{*}{-7.37e-06}	&\multirow{2}{*}{0.0045}	&\multirow{2}{*}{-0.6976}&\multirow{2}{*}{0.4884}
\\
\addlinespace
\multirow{2}{*}{pioneer}	& If any pioneer migrants within families =1; No=0	 & \multirow{2}{*}{1,128,624}	&\multirow{2}{*}{0.0237}	&\multirow{2}{*}{0.1522}	&\multirow{2}{*}{0}&\multirow{2}{*}{1}
\\
\bottomrule
\insertTableNotes 
\end{longtable}
}
\end{spacing}
\end{ThreePartTable}

\section{Estimation}

\subsection{Results}
\hspace{1.4em}Results in terms of Equations (15)--(20) are reported in Table 3. The most basic result in Column (1) indicates that a 10\% change in the ratio of job prospects (annual changes in sector-specific employment growth) at destination to origin is positively associated with an increase of 0.876 percentage points in migratory probabilities. It is worth noting that small $R^2$ values here are endemic to and a direct result of our discrete choice setting. Then, we added individual, household and city covariates to the second model, as seen in Column (2). The effects of job prospects increase considerably, from 0.876 to 1.933 percentage points, and additionally, having any pioneer migrants within families prior to the move leads to an increase of 6.67 percentage points in migratory probabilities. In addition, we estimate the same models using the distance in employment growth rates, i.e., $GR_{ik,t}-GR_{ij,t}$, and find that it has little impact on migration, as opposed to the trending indicator.\footnote{We show part of the results in \hyperref[tab4]{Table A3}.}\par

\hspace{1.4em}From Column (3) onward, we begin to eliminate unobserved heterogeneity where various options of fixed effects were added, and in Column (4), we included the CHRI (henceforth, the Hukou index). As reported in Column (3) and Columns (5)--(9), the magnitude of the coefficients of the ratio of sector-based job prospects in cities of destination to cities of origin is between 1.321 and 1.788 percentage points. Similarly, the magnitude of the coefficients of the family migration network is between 5.96 and 6.66 percentage points. Column (4) can be seen as a subpopulation analysis, because only 120 cities have Hukou indices. Results indicate that, after controlling for the Hukou registration stringency, the influence job prospects exert on migration climbs to 1.636 percentage points. And a unit increase in the distance in Hukou indices raises the probability of migration by 2.22 percentage points. The stringency gap \textit{per se} would not attract labour, but projects regional disparities in economic development and urbanisation.\footnote{The top four indexed cities are Beijing, Shanghai, Guangzhou, and Shenzhen, colloquially known as `Bei-Shang-Guang-Shen', representing the most developed areas in China.} However, the interaction effect between the Hukou index and job prospects is just statistically significant at the 10\% level.

\begin{sidewaystable}
\begin{ThreePartTable}
\begin{TableNotes}[flushleft]\footnotesize
\item Notes: Complete results of columns (1)--(9) are reported in Appendix A using the untransformed trending indicator. Standard errors shown in parentheses are clustered at the destination city.  \sym{*} \(p<0.10\), \sym{**} \(p<0.05\), \sym{***} \(p<0.01\)
\item Source: Created by authors using CHFS, CHRI and China Data Institute (2021).
\end{TableNotes}

\def\sym#1{\ifmmode^{#1}\else\(^{#1}\)\fi}
\fontsize{10}{12}\selectfont{
\begin{longtable}{l c c c c c c c c c}
\captionsetup{belowskip=0pt,aboveskip=1pt}
\caption{Determinants of Migration Decisions (1997--2017): Job Prospects with Fixed Effects}\\
\toprule\endfirsthead\midrule\endhead\midrule\endfoot\endlastfoot
\multicolumn{5}{l}{\textbf{\textit{No Restrictions}}} \\
\midrule
&\multicolumn{2}{c}{OLS}&\multicolumn{7}{c}{Multilateral Resistance to Migration} \\ \cmidrule(lr){2-3} \cmidrule(lr){4-10} 
                    &\multicolumn{1}{c}{(1)}  &\multicolumn{1}{c}{(2)}  &\multicolumn{1}{c}{(3)}  &\multicolumn{1}{c}{(4)}  &\multicolumn{1}{c}{(5)}  &\multicolumn{1}{c}{(6)}&\multicolumn{1}{c}{(7)}&\multicolumn{1}{c}{(8)}&\multicolumn{1}{c}{(9)}  \\
\midrule
distance\_JobTrend           &      0.0876\sym{***}&      0.1933\sym{***}&      0.1408\sym{***}&  0.1636\sym{***}&      0.1491\sym{***}&      0.1721\sym{***}&      0.1321\sym{***} &      0.1354\sym{***}&      0.1788\sym{***} \\
                    &    (0.0297)         &    (0.0290)         &    (0.0330)    & (   (0.0579)         &    (0.0329)         &    (0.0576)         &    (0.0321)     &    (0.0314)         &    (0.0576)     \\
distance\_CHRI        &                     &                     &                        &  0.0222\sym{**} &                     &                     &                     \\
                    &                     &                     &                    &    (0.0087)         &                     &                     &                     \\
\multicolumn{3}{l}{distance\_(JobTrendXCHRI)}                                          &       &               0.2212\sym{*}  &                     &                            \\
                    &                     &                     &           &            (0.1129)         &                     &                     &                     \\
pioneer          &                     &      0.0667\sym{***}&      0.0619\sym{***}    &         0.0576\sym{***}&      0.0666\sym{***}&      0.0596\sym{***}&      0.0614\sym{***} &      0.0617\sym{***}&      0.0601\sym{***} \\
                    &                     &    (0.0043)         &    (0.0035)    &    (0.0063)         &    (0.0045)         &    (0.0033)         &    (0.0036)     &    (0.0037)         &    (0.0034)      \\   
\addlinespace                    
Ind. Controls &   & Y  & Y & Y &  Y &  Y & Y & Y & Y \\
City Controls  &   &  Y & Y & Y & Y &  Y & Y & Y & Y \\
Constant           &      0.0090\sym{***}&      0.0622\sym{***}&      0.0603\sym{***}&    0.0406\sym{***}&      0.0396\sym{***}&      0.0607\sym{***}&      0.0600\sym{***}  &      0.0601\sym{***}&      0.0606\sym{***}\\
                    &    (0.0009)         &    (0.0045)         &    (0.0041)    &     (0.0038)         &    (0.0042)         &    (0.0042)         &    (0.0042)      &    (0.0042)         &    (0.0041)   \\
\addlinespace
Time FE & Y  & Y & Y  & Y  & Y &  &  \\
Industry FE &   &  & Y  & Y  & Y & Y & Y \\
Origin FE &   &   & Y &  Y &  &  Y & & &Y \\
Destination  FE &   &   & Y  & Y &  & & Y & Y &\\
Pairs of cities FE &  &   & & & Y & & \\
Origin-year FE & & & && & & Y & Y & \\
Dest-year FE & & & & && Y & & &Y\\
Industry-year FE & & & & &&  & & Y & Y \\
\midrule
$R^2$ & 0.0014       & 0.0459         & 0.0534       & 0.0509         & 0.0842         & 0.0721 & 0.0697    & 0.0696 & 0.0721    \\	
Obs                & 969998       & 749219         & 729965       & 408877         & 729960         & 729769 & 729725     & 748981         & 749034    \\ \bottomrule
\insertTableNotes 
\end{longtable}
}
\end{ThreePartTable}
\end{sidewaystable}

\FloatBarrier

\begin{ThreePartTable}
\setlength\LTleft{0pt}
\setlength\LTright{0pt}
\linespread{0.7}
\begin{TableNotes}[flushleft]\footnotesize
\item Notes: Here, regressions are conditional on the predicted probabilities produced by the models of Columns (3)--(7) reported in Table 3. Standard errors shown in parentheses are clustered at the destination city.  \sym{*} \(p<0.10\), \sym{**} \(p<0.05\), \sym{***} \(p<0.01\)
\item Source: Created by authors using CHFS, CHRI and China Data Institute (2021).
\end{TableNotes}

\scriptsize 
\setlength{\mycolwidth}{\dimexpr \textwidth/8 - 3\tabcolsep}
\def\sym#1{\ifmmode^{#1}\else\(^{#1}\)\fi}
\fontsize{10}{12}\selectfont{
\begin{longtable}{>{\raggedright\arraybackslash}p\mycolwidth C\mycolwidth C\mycolwidth C\mycolwidth C\mycolwidth C\mycolwidth C\mycolwidth C\mycolwidth C\mycolwidth}
\captionsetup{belowskip=0pt,aboveskip=1pt}
\caption{Determinants of Migration Decisions (1997--2017): Supplement}\\
\toprule\endfirsthead\midrule\endhead\midrule\endfoot\endlastfoot
\multicolumn{4}{l}{\textbf{\textit{Predicted Probabilities between 0 and 1}}} \\
\midrule
& &\multicolumn{1}{c}{(1)}  &\multicolumn{1}{c}{(2)} & \multicolumn{1}{c}{(3)}  &\multicolumn{1}{c}{(4)}& \multicolumn{1}{c}{(5)} & \multicolumn{1}{c}{(6)} & \multicolumn{1}{c}{(7)} \\
\midrule
distance\_JobTrend         &            &  0.1515\sym{***}& 0.2562\sym{**}&  0.1438\sym{***}    &  0.1947\sym{**}   &  0.1281\sym{***}   &      0.1312\sym{***}&      0.2185\sym{***} \\ &
                                     & (0.0454)     &  (0.1291)       &   (0.0480)        &  (0.0771)     &   (0.0431)      &    (0.0428)         &    (0.0776)  \\ 
distance\_CHRI           &              &    &  0.0292\sym{***} &                     &                     &                     \\ &
                          &               &   (0.0097)       &                     &                     &                     \\
distance\_(JobTrendXCHRI)                    &                                             & &    0.1933     &                     &                            \\
                                                  & & &             (0.1952)      &                     &                     &                     \\ 
pioneer       &          & 0.0625\sym{***}  &  0.0564\sym{***}     &  0.0666\sym{***}    &   0.0610\sym{***}    & 0.0626\sym{***}   &      0.0628\sym{***}&      0.0618\sym{***}  \\
                    &           & (0.0035)       &   (0.0059)        &   (0.0046)      &     (0.0033)      &      (0.0036) &    (0.0036)         &    (0.0033)    \\   
\addlinespace                    
\multicolumn{2}{l}{Ind. Controls}    & Y & Y & Y &  Y & Y & Y & Y \\
\multicolumn{2}{l}{City Controls}    & Y & Y & Y &  Y & Y  & Y & Y \\
Constant         &             &  0.0621\sym{***}  & 0.0431\sym{***} &  0.0386\sym{***}   &   0.0596\sym{***}   &  0.0614\sym{***}   &      0.0613\sym{***}&      0.0589\sym{***} \\
                       &            & (0.0042)      &    (0.0042)     &    (0.0041)       &    (0.0043)        &  (0.0042)     &    (0.0042)         &    (0.0042)   \\
\addlinespace
\multicolumn{2}{l}{Time FE}   & Y & Y & Y &    \\
\multicolumn{2}{l}{Industry FE}     & Y & Y & Y & Y & Y \\
\multicolumn{2}{l}{Origin FE}    & Y & Y &  &  Y & & & Y \\
\multicolumn{2}{l}{Destination  FE}     & Y & Y &  & & Y & Y \\
\multicolumn{2}{l}{Pairs of cities FE}     & & & Y & & \\
\multicolumn{2}{l}{Origin-year FE}  & && & & Y & Y &\\
\multicolumn{2}{l}{Dest-year FE}   & & && Y & && Y \\
\multicolumn{2}{l}{industry-year FE}   & & && & &Y & Y \\ 
\midrule
$R^2$    &    & 0.0564         & 0.0558       & 0.0878         & 0.0801         & 0.0750    & 0.0748 & 0.0804     \\	
Obs           &                 &668644     &   306227   &   724519     &  680253 &     678702   & 696327        & 701926     \\
\bottomrule
\insertTableNotes 
\end{longtable}
}
\end{ThreePartTable}

\hspace{1.4em}In Table 4, we restricted observations to those whose predicted probabilities are within the unity. Except for Column (2), all regressions involved more than 90\% of observations. Estimates of job prospects in Columns (1), (2), (4) and (7) are found to be larger than their counterparts reported in Columns (3), (4), (6) and (9) of Table 3. In contrast, Columns (3), (5) and (6) show slightly smaller effects of job prospects on migration decisions compared to Columns (5), (7) and (8) of Table 3. In sum, the magnitude of the coefficients of the ratio of sector-based job prospects at destination to origin is between 1.281 and 2.185 percentage points. Furthermore, the interaction term is statistically insignificant. Given these results, we conclude that our linear probability estimator performs acceptably.\footnote{After computing the predicted probabilities of each model, we counted the proportion of migrant observations with negative fitted values. The proportions are 1.1\%, 0.9\%, 0.7\%, 2.4\%, 1.4\%, 1.4\%, and 2.4\% of the total.}\par

\hspace{1.4em}Regarding estimates of control variables reported in \hyperref[tab2]{Table A1}, we see that migratory probabilities increase for people who are male, unmarried, younger, or more educated. It is noteworthy that \textcite{He2003} point out that migrant men were predominantly driven by economic incentives, while women moved chiefly for social or family reasons. Because we focus on labour migrants, our findings are distinguished from those studies that do not primarily analyse labour migration. For example, in this study, men are found to be more likely to migrate than women, whereas an opposite storey is not uncommon among other migration scenarios, such as permanent migration (e.g., \cites{Meng2020, Zhang2020}). As for city covariates, the distance in income is found statistically insignificant in Columns (3)--(9) when we further control for heterogeneity in migration preferences. Instead, two variables are consistently significant, that is, the distance in the provision of healthcare and business density. As expected, abundant medical resources positively drive migration, while the negative role of business density in migration seems surprising.\footnote{With a simple regression, the sign remains negative.}\par

\subsection{Robustness Check}

\subsubsection{Multilevel Logistic Regression}
\hspace{1.4em}Results of multilevel logit models are reported in Columns (1)--(6) of Table 5. In Column (1), we consider cities of origin as the higher level to control for origin-specific factors affecting the probability of migration. As we see, the intra-class correlation coefficient (ICC) is merely 0.0432, indicating that observations within the same place of origin are different from each other, so we switched to Columns (2) and (3) where cities of destination and origin-destination pairs are treated as level-2, respectively.\footnote{ICC is calculated as the ratio of the between-group variance relative to the total variance in the sample. It describes the extent to which observations within city groups are similar to each other.} Here, the ICC is 0.3355 for destination and 0.2783 for pairwise cities, presenting substantial evidence of clustering, and their coefficients are smaller than in Column (1). In other words, observations within the same destination (origin-destination pairs) have a much lower degree of variability compared to their origin-nested counterparts.\footnote{This is the reason, in addition to the CHFS's sampling design, for us to cluster standard errors at the destination city for all non-hierarchical models (see \cite{ColinCameron2015}).} As destination-nested models present the largest ICC, we mainly interpret these results.\footnote{The pairwise nested model produces results most similar to the fixed-effects models but has a lower ICC than the destination-nested model.} Because coefficients in the logistic regression are not as easily interpreted as coefficients in the linear regression, we graphically illustrate marginal effects of the second model with an interval of 0.2 in Figure 5a, holding all other variables at mean. The plot clearly shows an increasing trend that, when the ratio of job prospects at destination to origin gets larger, its effects on motivating migration become increasingly stronger.\par

\hspace{1.4em}It has been widely demonstrated that education levels have a considerable impact on the propensity for migration and location preferences (e.g., \cites{Meng2020, Fu2012}). Thus, we treated years of schooling as a new level, subordinate to cities. In other words, people with identical educational attainments nested in the same locations are supposedly much more similar to each other than their fellow migrants who received a higher or lower level of education. As can be seen, the ICC becomes slightly larger in Columns (4)--(5) than in Columns (1)--(2), whereas almost no change arises from adding the level of education under pairwise nests. It should be noted that cities are level-2 in Columns (1)--(3), but level-3 in Columns (3)--(6). Hence, the variance between cities is 0.1484, 1.6611, and 1.2684 in two-level logit models and 0.1054, 1.6046, and 1.2685 in three-level logit models. Interestingly, the distinction between education groups is even greater than the extent to which that the grouping of origin can account for, as opposed to Column (6), where the variance between level-2 groups within the same origin-destination pairs is nearly zero. The fifth model gives an in-between result: its level-2 variance is around one-seventh of the level-3. We further plot its marginal effects, holding all covariates at mean. As seen in Figure 5b, the marginal effects are also continuously increasing.\par

\vspace{0.5cm}
\renewcommand\arraystretch{0.5}
\begin{ThreePartTable}
\linespread{0.7}
\begin{TableNotes}[flushleft]\footnotesize
\item Notes: In Appendix A, we report complete results using the untransformed trending indicator. AR(2) is the Arellano-Bond test for second-order serial correlation with the null hypothesis of no serial correlation in disturbances. Hansen's J test is a test of over-identifying restrictions (in other words, the overall validity of the instruments) using J statistic of \textcite{Hansen1982}. Likewise, the difference-in-Hansen test is designed to test the validity of subsets of the instruments. Endogenous and predetermined variables are instrumented with their corresponding second- and third-order lags in column (5)--(6). Three more orders of lags are added in column (7). The trending indicator, the income (GDP per capita), and the share of the tertiary sector are treated as endogenous in column (5), while other city-level covariates are treated as predetermined (not strictly exogenous). In column (6)--(7), the population density and business density are additionally treated as endogenous. Moreover, individual and household covariates are specified as exogenous. The Windmeijer correction \autocite{Windmeijer2005} is used in the GMM estimation, and corresponding standard errors are clustered at the destination city. Besides, robust standard errors are also applied in multilevel logit models.\sym{*} \(p<0.10\), \sym{**} \(p<0.05\), \sym{***} \(p<0.01\)
\item Source: Created by authors using CHFS and China Data Institute (2021).
\end{TableNotes}

\scriptsize 
\setlength{\mycolwidth}{\dimexpr \textwidth/10 - 0.75\tabcolsep}
\def\sym#1{\ifmmode^{#1}\else\(^{#1}\)\fi}
\setlength{\tabcolsep}{2pt}
\fontsize{9}{12}\selectfont{
\begingroup
\setlength{\LTleft}{-20cm plus -1fill}
\setlength{\LTright}{\LTleft}
\begin{longtable}{>{\raggedright\arraybackslash}p\mycolwidth C\mycolwidth C\mycolwidth C\mycolwidth C\mycolwidth C\mycolwidth C\mycolwidth C\mycolwidth C\mycolwidth C\mycolwidth C\mycolwidth}
\captionsetup{belowskip=0pt,aboveskip=1pt}
\caption{Determinants of Migration Decisions: Multilevel Logit and Two-step System GMM}\\
\toprule\endfirsthead\midrule\endhead\midrule\endfoot\endlastfoot
&\multicolumn{3}{c}{Two Level Logit}&\multicolumn{3}{c}{Three Level Logit}&\multicolumn{3}{c}{GMM} \\ \cmidrule(lr){2-4} \cmidrule(lr){5-7} \cmidrule(lr){8-10}
                    &\multicolumn{1}{c}{(1)}  &\multicolumn{1}{c}{(2)}  &\multicolumn{1}{c}{(3)}  &\multicolumn{1}{c}{(4)} & \multicolumn{1}{c}{(5)}  &\multicolumn{1}{c}{(6)}& \multicolumn{1}{c}{(7)} & \multicolumn{1}{c}{(8)} & \multicolumn{1}{c}{(9)} \\
\midrule
distance\_ JobTrend           &      5.3903\sym{***}&      4.5964\sym{***}& 3.3363\sym{***}  &   5.3922\sym{***}&      4.5847\sym{***}&   3.3363\sym{***}     &0.3157\sym{***}&      0.2493\sym{***}&      0.2751\sym{***}\\
                    &    (1.0279)         &    (0.6027)         & (1.0040)   &   (1.0181)         &    (0.6120)    & (1.0040)     &   (0.0754)         &    (0.0569)         &    (0.0582)         \\
\addlinespace                    
Ind. Controls & Y   & Y  & Y  & Y &Y &Y &  Y & Y & Y \\
City Controls  &  Y &  Y & Y  & Y & Y& Y &  Y & Y & Y \\
Intercept            &     -1.839\sym{***}&     -1.5841\sym{***}&  0.0077   &  -1.9504\sym{***}&     -1.7681\sym{***}&   0.0077  & 0.0610\sym{***}&      0.0590\sym{***}&      0.0590\sym{***}\\
                    &    (0.3101)         &    (0.2573)         &  (0.2468)  &    (0.3137)         &    (0.2857)     &  (0.2468)   &    (0.0046)         &    (0.0046)         &    (0.0046)         \\
\addlinespace
Level 2 var.    &    0.1484                 &    1.6611            &   1.2684  &      0.1869 &  0.2416       &    1.99e-33        &                     &                     &                     \\
                    &              (0.0372)         &    (0.1671)        &  (0.1009)        &  (0.0312)  &     (0.0455)      &  (3.38e-34)      &                     &                     &                     \\
Level 3 var.    &                    &             &   &      0.1054 &   1.6046       &  1.2685        &                   &                     &                     \\ &                   &               &      &  (0.0398)   &    (0.1685)      &  (0.1009)    &                     &                     &                     \\
ICC  &            0.0432                &    0.3355            &     0.2783  &   0.0816  &    0.3595       &    0.2783  &                     &                     &                   \\
&            (0.0104)                &    (0.0224)         &    (0.0160)       &  (0.0112)   &       (0.0214)   & (0.0160)    &                     &                     &                   \\
\addlinespace
Nest & origin & destination & pair & origin & destination & pair \\
sub-Nest  & & & & education & education & education \\
\addlinespace
Time FE &             Y          &        Y             &          Y             &         Y         &           Y              &                    Y   &     Y      &     Y     &     Y       \\
Industry FE &                     &             &       &                      &                      &    &     Y              &                    Y   &     Y                 \\
\addlinespace
\multicolumn{2}{l}{Num. of instruments}            &          &                 &    &                     &            &            182                &                    180 &     201                \\
AR(2) &                     &                     &        &             &   &                  &    0.724                &                    0.790 &   0.790               \\
\multicolumn{2}{l}{Hansen's J test}                     &          &           &                 &          &           &      0.349               &                    0.263 &       0.404              \\
\addlinespace
\multicolumn{3}{l}{\textit{Difference-in-Hansen tests}}   \\
\multicolumn{5}{l}{GMM instruments for levels -- Excluding group}                 &           &              & 0.114 & 0.108 & 0.154 \\
\multicolumn{6}{l}{GMM instruments for levels -- Difference (null H = exogenous)}                                           & & 0.451 & 0.355 & 0.578 \\
\multicolumn{6}{l}{GMM instrument for distance\_trend -- Excluding group}        &         &0.520 & 0.354 & 0.526 \\
\multicolumn{7}{l}{GMM instrument for distance\_trend -- Difference (null H = exogenous)}       &0.114 & 0.192 & 0.173 \\
\midrule
Obs                 & 749219        & 749219     &  749219  & 749219         & 749219  &   749219  &    729965        & 729965 & 729965         \\
\bottomrule
\insertTableNotes 
\end{longtable}
\endgroup
}
\end{ThreePartTable}
\FloatBarrier

\hspace{1.4em}In \hyperref[tab3]{Table A2}, we report the complete regression results of Table 5. The results of covariates vary depending on the definition of nests, except for the distance in income, which is consistently statistically significant across all models. The distance in the provision of healthcare (business density) is found positive (negative) at the 5\% (10\%) level in pairwise nested models but not in others. In contrast, the distance in population density and the share of the tertiary sector exhibit positive impacts on increasing the probability of migration.\footnote{The relationship between the distance in the share of the tertiary sector and migration decisions is negative when we account for origin- and destination-specific effects simultaneously as in pairwise nested models, despite being statistically insignificant.}\par

\begin{figure}[!tbp]
  \begin{subfigure}[b]{0.5\textwidth}
  \centering
    \includegraphics[scale=0.55]{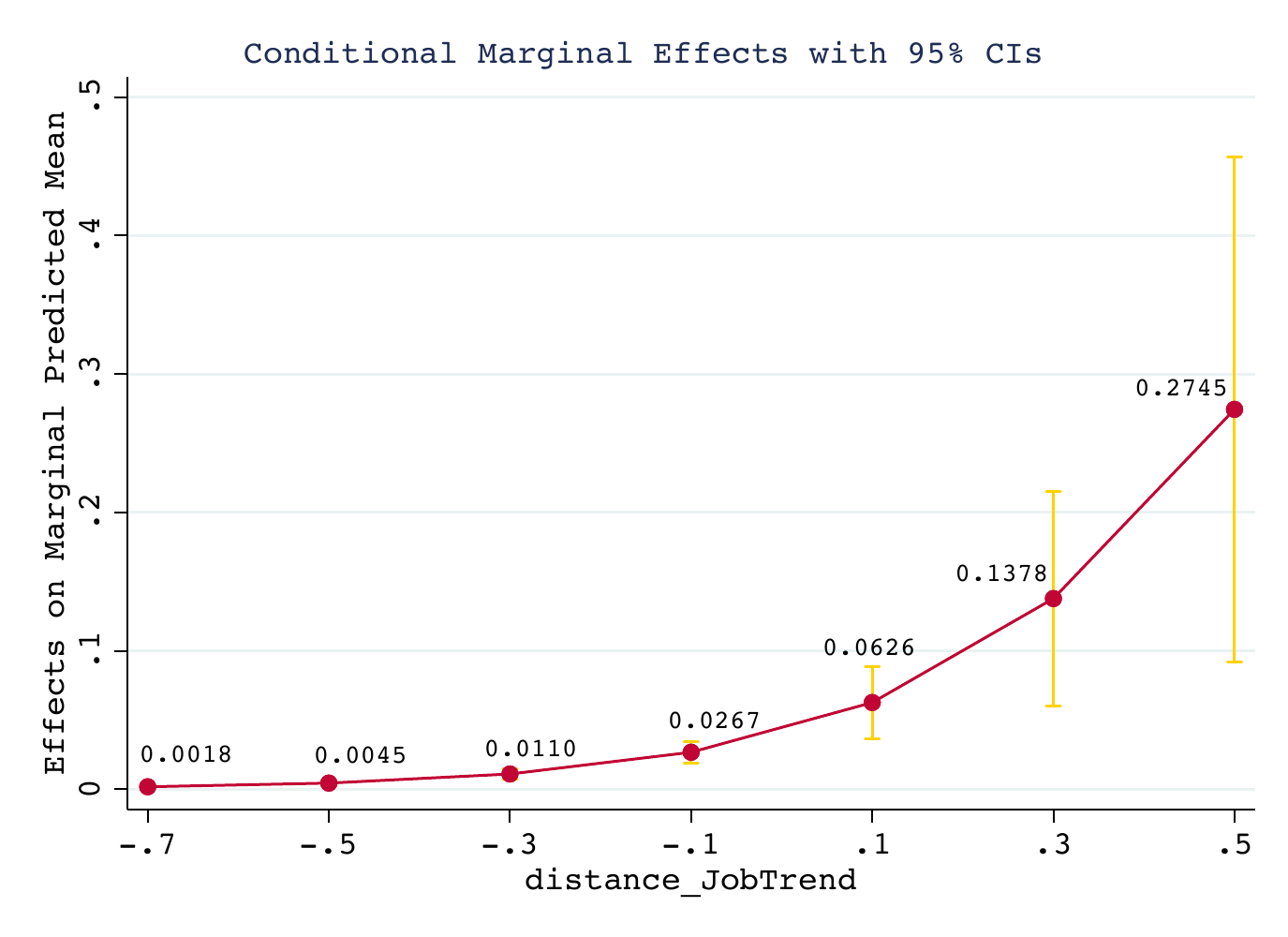}
    \caption{\fontsize{10}{12}\selectfont{Two-level random intercept model.}}
    \label{fig:f1}
  \end{subfigure}
  \hfill
  \begin{subfigure}[b]{0.5\textwidth}
  \centering
    \includegraphics[scale=0.55]{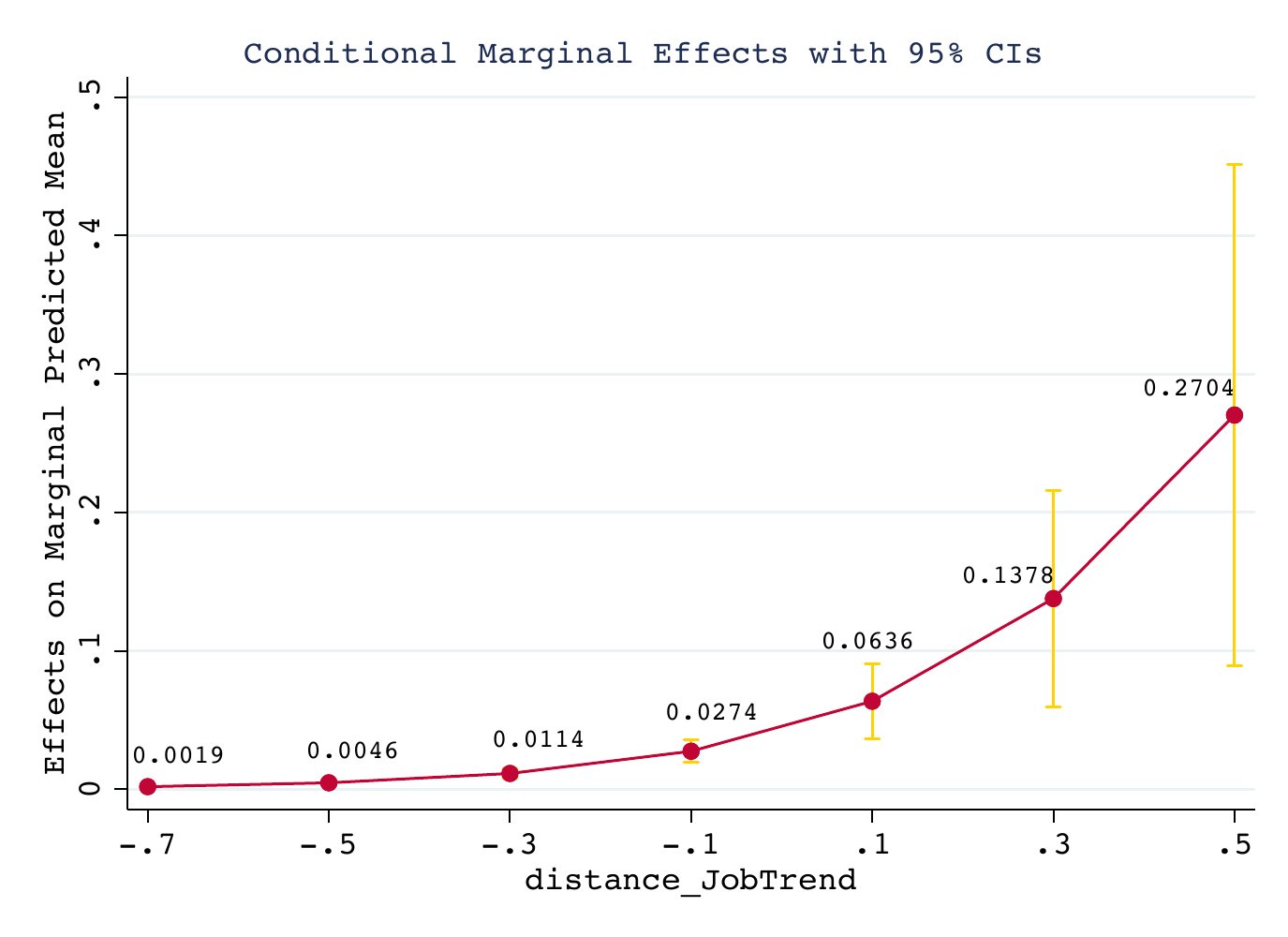}
    \caption{\fontsize{10}{12}\selectfont{Three-level random intercept model.}}
    \label{fig:f2}
  \end{subfigure}
  \caption{\fontsize{10}{12}\selectfont{Marginal effects of destination-nested model. \\Notes: All results are statistically significant at the 1\% level.} \\ Source: Created by authors using CHFS and China Data Institute (2021).}
\end{figure}
\FloatBarrier

\subsubsection{Two-step System GMM}
\hspace{1.4em}In Columns (7)--(9) of Table 5, we report two-step system GMM estimates with different lag and instrument strategies. We adopted two types of fixed effects: time and industry. The former is aimed at absorbing any instant shock exposed to all units, while the latter is concerned to ensure that variations of job prospects are estimated within sectors. The first GMM model is the baseline where we utilised two orders of lags to instrument the endogenous and predetermined variables. We initially treated the trending indicator, the income, and the share of the tertiary sector as strictly endogenous, while all other city covariates were treated as predetermined. The coefficient of the trending indicator shows that a 10\% change in the ratio of sector-based job prospects at destination to origin causes an increase of 3.157 percentage points in migratory probabilities. Then, we applied the same lag strategy but, additionally, treated the population density and business density as endogenous variables. As a result, the coefficient is still statistically significant at the 1\% level but relatively smaller than the previous result. We kept this instrument strategy but added three more orders of lags to instrument our variables. Adding higher orders of lags can help remove serial correlation but may also impose the weak instrument problem, so we did not include more lags. We learn from Column (9) that the magnitude of the coefficient of our trending indicator is a bit larger but falls exactly between the values of coefficients in Columns (7) and (8). Moreover, the results here are larger than the fixed effects estimates reported in Table 3 and, interestingly, the model of Column (9), where we included the Hukou index and confined observations to having a within-unity predicted probability, produces the closest estimate (2.562 percentage points in Table 4).\par

\hspace{1.4em}As seen in \hyperref[tab3]{Table A2}, half of the city-level control variables are still statistically insignificant in Column (7). While in Columns (8)--(9), the majority appears to be statistically significant, as opposed to the business density, which was initially found to have an impact, but then became negligible when it was treated as endogenous. The second and third GMM models are better than the first in dealing with unobserved heterogeneity, their results are thus more reliable. According to these results, we find that the distance in income, the provision of healthcare, the provision of higher education and population density have positive effects on driving migration. In contrast, the distance in the share of the tertiary sector negatively affects migratory probabilities. The relationship between the distance in population density and migration uncovered here can be attributed to the fact that migrants are attracted to large cities that inevitably are densely populated \autocite{Chen2016}. For the latter, a possible explanation is that considering, in our sample, 39\% of total migrants worked in the tertiary sector, cities with a relatively lower share of the tertiary sector are more likely to present a faster pace of growth.\footnote{Based on the variance inflation factor, the colinearity between the trending indicator and the share of the tertiary sector is very weak (their VIFs are 1.01 and 3.01, respectively). Also, the Pearson's correlation coefficient is just 0.0021.}\par

\hspace{1.4em}The results of Hansen's J test for over-identification are well above 0.25, a threshold suggested by \textcite{Roodman2009}, but far from 1, pointing to a 34.9\%, 26.3\% and 40.4\% chance of a type one error if the null is rejected, and no issue of instrument proliferation. The null hypothesis of the Arellano--Bond test is not rejected, indicating no second-order serial correlation in disturbances.\footnote{Despite not being reported, no serial correlation was found in AR(3)--(4).} By further checking the difference-in-Hansen test for the validity of subsets of instruments, the differenced models are evidenced as dynamically complete, implying the instruments used in the level models are valid. We also report test results of GMM instruments of the trending indicator, from which we can conclude that its corresponding specified instruments used in the level models are exogenous.\par

\section{Discussions and Concluding Remarks}

\hspace{1.4em}In this paper, we investigated the effects of sector-based job prospects on individual migration decisions across Chinese prefecture boundaries. To this end, we assembled a unique quasi-panel based on the 2017 China Household Finance Survey and the prefecture city statistics between 1997 and 2017. By accounting for city-level bilateral variations in parallel with individual and household characteristics, we filled gaps in the existing Chinese migration literature in two aspects: a) migration decisions at the city level are quite scarce due to data limitations, and among them, individual and household characteristics are always absent in analysing regional longitudinal effects, and b) the previous models that have controlled for these characteristics rely on micro-data whereby regional factors are either missing, monadic, or at the province level. Equally importantly, Chinese migration research is desperately lacking in visionary migration scenarios. By constructing a proxy variable for job prospects, we enriched economic incentives of labour migration from a forward-looking angle. Further, we theoretically added to the literature of migration decision-making by synthesising the virtues of dynamic discrete choice models and reference-dependence derived from prospect theory. Following it, we drew corresponding empirical specifications and applied various monadic and dyadic fixed effects to address multilateral resistance to migration. Besides, we considered multilevel logistic regression and two-step system GMM estimation for the robustness check. In sum, this study, acclimatized to the new migration pattern that individuals tend to move within provinces, deepened the understanding of relationships between regional employment structures and labour mobility in both level and scope.\par

\hspace{1.4em}Our results primarily indicate that a 10\% increase in the ratio of job prospects in cities of destination to cities of origin raises the probability of migration by 1.281--2.185 percentage points, and the effects tend to be stronger when the scale of the ratio is larger. Having a family migration network causes an increase of approximately 6 percentage points in migratory probabilities. Additionally, labour migrants are more likely to be male, unmarried, younger, or more educated. Here, the message our findings deliver is rather simple -- for medium- and small-sized cities that are less competent in retaining or attracting labour, promoting a promising labour market could help, and downward migration may even happen in this case. Future research could pay attention to fieldwork to learn more personal, subjective beliefs from labour migrants to enrich the evaluative dimension of job prospects.\par

\hspace{1.4em}China's extensive, rapid development of infrastructure and transportation boosts the circulation of human capital and labour forces between regions. A high-speed rail now takes only five hours to complete 1318 kilometres between Shanghai and Beijing. These factors underlie the persistence of high labour mobility into the future. At the same time, migrants were found to be increasingly older and more educated from 2000 to 2015 \autocite{Commission2018}. Significantly, only 2\% of migrants had at least a college degree in 1990, while 25 years later, the percentage is 23.3\%. Skilled migrants tend to prioritise career prospects over quality of life in the migration decision-making process \autocite{Liu2014}. This trend well signifies why we should re-examine economic incentives from the angle of job prospects. It is evident that China's current and future cohorts, who are more educated, are more likely to make visionary migration decisions than their elder generations. The likelihood of getting a job and how good it is are the most common questions with which (potential) workers are concerned. A variety of dimensions can contribute to assessing job quality, and what we have attempted to highlight in this paper is but one. Individual fulfilment and aspirations that are much more personal and sophisticated, as \textcite{Becker2020} discuss, can also affect how people regard their job prospects across regions. Our point hereby is basic and general.\par


\clearpage 

\begingroup
\setstretch{0.9}
\setlength\bibitemsep{5pt}
\AtNextBibliography{\small}
\printbibliography
\endgroup

\section*{Appendix A: Regression Tables}
\setcounter{table}{0}
\setcounter{figure}{0}
\renewcommand{\thetable}{A\arabic{table}}
\renewcommand{\thefigure}{A\arabic{figure}}

\begin{sidewaystable}
\begin{ThreePartTable}
\linespread{0.7}
\begin{TableNotes}[flushleft]\footnotesize
\item Notes: Trending indicators estimated here are untransformed, i.e., $Job\_Trending_{ik,t}-Job\_Trending_{ij,t}$. Standard errors shown in parentheses are clustered at the destination city. Coefficients are displayed as 0.0000 because the values are smaller than 0.0001. The small $R^2$  is endemic to and a direct result of our discrete choice setting. \sym{*} \(p<0.10\), \sym{**} \(p<0.05\), \sym{***} \(p<0.01\)
\item Source: Created by authors using CHFS, CHRI and China Data Institute (2021).
\end{TableNotes}

\def\sym#1{\ifmmode^{#1}\else\(^{#1}\)\fi}
\fontsize{9}{12}\selectfont{
\setlength{\tabcolsep}{6pt}
\begin{longtable}{l  c c c c c c c c c}
\captionsetup{belowskip=0pt,aboveskip=1pt}
\caption{Determinants of Migration Decisions (1997--2017): Job Prospects with Fixed Effects\label{tab2}}\\
\toprule\endfirsthead\midrule\endhead\midrule\endfoot\endlastfoot
&\multicolumn{2}{c}{OLS}&\multicolumn{7}{c}{Multilateral Resistance to Migration} \\ \cmidrule(lr){2-3} \cmidrule(lr){4-10} 
                    &\multicolumn{1}{c}{(1)}  &\multicolumn{1}{c}{(2)}  &\multicolumn{1}{c}{(3)}  &\multicolumn{1}{c}{(4)}  &\multicolumn{1}{c}{(5)}  &\multicolumn{1}{c}{(6)}&\multicolumn{1}{c}{(7)}&\multicolumn{1}{c}{(8)}&\multicolumn{1}{c}{(9)}  \\
\midrule
distance\_JobTrend           &      0.0011\sym{***}&      0.0020\sym{***}&      0.0014\sym{***}&      0.0018\sym{***}&      0.0015\sym{***}&      0.0017\sym{***}&      0.0013\sym{***}  &      0.0013\sym{***}&      0.0018\sym{***}\\
                    &    (0.0003)         &    (0.0003)         &    (0.0003)         &    (0.0007)         &    (0.0003)         &    (0.0005)         &    (0.0003)     &    (0.0003)         &    (0.0005)     \\
gender            &                     &      0.0009\sym{***}&      0.0006\sym{***}&      0.0005\sym{***}&      0.0002         &      0.0006\sym{***}&      0.0006\sym{***} &      0.0006\sym{***}&      0.0006\sym{***}\\
                    &                     &    (0.0001)         &    (0.0001)         &    (0.0002)         &    (0.0001)         &    (0.0001)         &    (0.0001)        &    (0.0001)         &    (0.0001) \\
marriage            &                     &     -0.0080\sym{***}&     -0.0074\sym{***}&     -0.0048\sym{***}&     -0.0041\sym{***}&     -0.0072\sym{***}&     -0.0074\sym{***} &     -0.0074\sym{***}&     -0.0073\sym{***} \\
                    &                     &    (0.0006)         &    (0.0005)         &    (0.0006)         &    (0.0005)         &    (0.0005)         &    (0.0005)     &    (0.0005)         &    (0.0005)    \\
hukou\_type          &                     &     -0.0010\sym{**} &     -0.0004         &     -0.0008\sym{**} &      0.0000         &     -0.0004         &     -0.0004     &     -0.0004         &     -0.0004     \\
                    &                     &    (0.0004)         &    (0.0003)         &    (0.0003)         &    (0.0002)         &    (0.0003)         &    (0.0003)      &    (0.0003)         &    (0.0003)    \\
health\_status       &                     &     -0.0002\sym{**} &     -0.0001         &      0.0000         &     -0.0001\sym{*}  &     -0.0001\sym{*}  &     -0.0001    &     -0.0001\sym{*}  &     -0.0001\sym{*}     \\
                    &                     &    (0.0001)         &    (0.0001)         &    (0.0001)         &    (0.0001)         &    (0.0001)         &    (0.0001)   &    (0.0001)         &    (0.0001)        \\
hh\_income &                     &      0.0000         &     -0.0000         &      0.0001         &     -0.0001\sym{**} &      0.0000         &     -0.0000    &   -0.0000         &      0.0000  \\
                    &                     &    (0.0001)         &    (0.0001)         &    (0.0001)         &    (0.0001)         &    (0.0001)         &    (0.0001)      &    (0.0001)         &    (0.0001)     \\
age                 &                     &     -0.0025\sym{***}&     -0.0023\sym{***}&     -0.0016\sym{***}&     -0.0014\sym{***}&     -0.0023\sym{***}&     -0.0023\sym{***}&     -0.0023\sym{***}&     -0.0023\sym{***} \\
                    &                     &    (0.0002)         &    (0.0002)         &    (0.0002)         &    (0.0002)         &    (0.0002)         &    (0.0002)    &    (0.0002)         &    (0.0002)     \\
age$^2$    &                     &      0.0000\sym{***}&      0.0000\sym{***}&      0.0000\sym{***}&      0.0000\sym{***}&      0.0000\sym{***}&      0.0000\sym{***} &      0.0000\sym{***}&      0.0000\sym{***} \\
                    &                     &    (0.0000)         &    (0.0000)         &    (0.0000)         &    (0.0000)         &    (0.0000)         &    (0.0000)      &    (0.0000)         &    (0.0000)     \\
schooling           &                     &      0.0002\sym{***}&      0.0002\sym{***}&      0.0002\sym{***}&      0.0000         &      0.0002\sym{***}&      0.0002\sym{***} &      0.0002\sym{***}&      0.0002\sym{***}\\
                    &                     &    (0.0000)         &    (0.0000)         &    (0.0000)         &    (0.0000)         &    (0.0000)         &    (0.0000)    &    (0.0000)         &    (0.0000)      \\
pioneer          &                     &      0.0667\sym{***}&      0.0619\sym{***}&      0.0575\sym{***}&      0.0666\sym{***}&      0.0596\sym{***}&      0.0614\sym{***} &      0.0617\sym{***}&      0.0601\sym{***} \\
                    &                     &    (0.0043)         &    (0.0035)         &    (0.0063)         &    (0.0045)         &    (0.0033)         &    (0.0036)      &    (0.0037)         &    (0.0034)     \\
distance\_Ingdppc       &                     &      0.0144\sym{***}&      0.0003         &      0.0034         &      0.0038         &     -0.0019         &     -0.0016  &     -0.0020         &     -0.0027        \\
                    &                     &    (0.0023)         &    (0.0044)         &    (0.0062)         &    (0.0045)         &    (0.0055)         &    (0.0045)    &    (0.0045)         &    (0.0054)      \\
distance\_coop &                     &     -0.0004         &     -0.0007\sym{**} &     -0.0003         &     -0.0007\sym{**} &     -0.0008\sym{**} &     -0.0008\sym{***} &     -0.0008\sym{***}&     -0.0008\sym{*}\\
                    &                     &    (0.0003)         &    (0.0003)         &    (0.0003)         &    (0.0003)         &    (0.0004)         &    (0.0003)    &    (0.0003)         &    (0.0004)      \\
distance\_medical    &                     &      2.2892\sym{**} &      3.2870\sym{***}&      2.4882\sym{**} &      3.0634\sym{***}&      4.1929\sym{***}&      3.2512\sym{***}  &      3.2223\sym{***}&      4.1993\sym{***}\\
                    &                     &    (1.0015)         &    (0.9410)         &    (0.9847)         &    (0.8152)         &    (1.2921)         &    (1.0355)     &    (1.0390)         &    (1.2887)     \\
distance\_highEdu&                     &      0.1441\sym{*}  &      0.0961         &     -0.0415         &      0.0404         &      0.1681\sym{*}  &      0.1215   &      0.1262         &      0.1716\sym{*}       \\
                    &                     &    (0.0812)         &    (0.0922)         &    (0.1496)         &    (0.0871)         &    (0.0941)         &    (0.0948)     &    (0.0903)         &    (0.0886)      \\
distance\_ppDen       &                     &      0.0026         &     -0.0016         &     -0.0008         &      0.0008         &     -0.0097         &     -0.0007   &     -0.0006         &     -0.0096       \\
                    &                     &    (0.0019)         &    (0.0084)         &    (0.0055)         &    (0.0074)         &    (0.0075)         &    (0.0080)     &    (0.0080)         &    (0.0074)     \\
\multicolumn{2}{l}{distance\_tertiaryRatio}                    &      0.0024         &     -0.0348         &     -0.0259         &     -0.0380         &     -0.0306         &     -0.0335   &     -0.0368         &     -0.0320      \\
                    &                     &    (0.0188)         &    (0.0254)         &    (0.0342)         &    (0.0249)         &    (0.0273)         &    (0.0253)    &    (0.0252)         &    (0.0271)      \\
distance\_CHRI         &                     &                     &                     &      0.0221\sym{**}&                     &                     &                     \\
                    &                     &                     &                     &    (0.0088)         &                     &                     &                     \\
\multicolumn{3}{l}{distance\_JobTrendXdistance\_CHRI}                                         &                     &     0.0022\sym{**}&                     &                     &                     \\
                    &                     &                     &                     &    (0.0010)         &                     &                     &                     \\
Constant            &      0.0089\sym{***}&      0.0622\sym{***}&      0.0604\sym{***}&      0.0405\sym{***}&      0.0396\sym{***}&      0.0607\sym{***}&      0.0600\sym{***} &      0.0601\sym{***}&      0.0606\sym{***}\\
                    &    (0.0009)         &    (0.0045)         &    (0.0041)         &    (0.0038)         &    (0.0042)         &    (0.0042)         &    (0.0042)     &    (0.0042)         &    (0.0041)      \\
\midrule
Time FE & Y  & Y & Y  & Y  & Y &  &  \\
Industry FE &   &  & Y  & Y  & Y & Y & Y \\
Origin FE &   &   & Y &  Y &  &  Y & & &Y \\
Destination  FE &   &   & Y  & Y &  & & Y & Y &\\
Pairs of cities FE &  &   & & & Y & & \\
Origin-year FE & & & && & & Y & Y & \\
Dest-year FE & & & & && Y & & &Y\\
Industry-year FE & & & & &&  & & Y & Y \\
$R^2$ & 0.0014       & 0.0459         & 0.0534       & 0.0509         & 0.0842         & 0.0721 & 0.0697    & 0.0696 & 0.0721    \\	
Obs                 & 969998         & 749219         & 729965         & 404487         & 729960         & 729769         & 729725    & 748981       & 749034      \\
\bottomrule
\insertTableNotes 
\end{longtable}
}
\end{ThreePartTable}
\end{sidewaystable}

\begin{ThreePartTable}
\linespread{0.7}
\begin{TableNotes}[flushleft]\footnotesize
\item Notes: Trending indicators estimated here are untransformed, i.e., $Job\_Trending_{ik,t}-Job\_Trending_{ij,t}$. Standard errors shown in parentheses are clustered at the destination city, except for multilevel logit models, where robust standard errors are applied but not clustered. The Windmeijer correction \autocite{Windmeijer2005} is enabled in the GMM estimation. Coefficients are displayed as 0.0000 because the values are smaller than 0.0001. \sym{*} \(p<0.10\), \sym{**} \(p<0.05\), \sym{***} \(p<0.01\)
\item Source: Created by authors using CHFS and China Data Institute (2021).
\end{TableNotes}
\newpage
\def\sym#1{\ifmmode^{#1}\else\(^{#1}\)\fi}
\fontsize{9}{12}\selectfont{
\setlength{\tabcolsep}{1pt}
\begin{longtable}{l c c c c c c c c c}
\captionsetup{belowskip=0pt,aboveskip=1pt}
\caption{Determinants of Migration Decisions: Multilevel Logit and Two-step System GMM\label{tab3}}\\
\toprule\endfirsthead\midrule\endhead\midrule\endfoot\endlastfoot
&\multicolumn{3}{c}{Two Level Logit}&\multicolumn{3}{c}{Three Level Logit}&\multicolumn{3}{c}{GMM} \\ \cmidrule(lr){2-4} \cmidrule(lr){5-7} \cmidrule(lr){8-10}
                    &\multicolumn{1}{c}{(1)}  &\multicolumn{1}{c}{(2)}  &\multicolumn{1}{c}{(3)}  &\multicolumn{1}{c}{(4)} & \multicolumn{1}{c}{(5)}  &\multicolumn{1}{c}{(6)}& \multicolumn{1}{c}{(7)} & \multicolumn{1}{c}{(8)} & \multicolumn{1}{c}{(9)} \\
\midrule
distance\_ JobTrend           &      0.0533\sym{***}&      0.0446\sym{***}&      0.0325\sym{***}&      0.0534\sym{***}&      0.0445\sym{***}&      0.0325\sym{***} &      0.3157\sym{***}&      0.2493\sym{***}&      0.2751\sym{***}\\
                    &   (0.0106)         &    (0.0058)         &    (0.0104)         &    (0.0106)         &    (0.0058)         &    (0.0104)      &    (0.0754)         &    (0.0569)         &    (0.0582)         \\
gender           &      0.1153\sym{***}&      0.0808\sym{***}&      0.0303\sym{*}  &      0.1265\sym{***}&      0.0950\sym{***}&      0.0303\sym{*} &                    0.0010\sym{***}&      0.0010\sym{***}&      0.0010\sym{***}\\
                    &    (0.0238)         &    (0.0192)         &    (0.0159)         &    (0.0243)         &    (0.0194)         &    (0.0159)     &    (0.0002)         &    (0.0002)         &    (0.0002)     \\
marriage            &       -0.4273\sym{***}&     -0.3765\sym{***}&     -0.4047\sym{***}&     -0.4110\sym{***}&     -0.3520\sym{***}&     -0.4047\sym{***} &     -0.0078\sym{***}&     -0.0076\sym{***}&     -0.0076\sym{***}\\
                    &   (0.0302)         &    (0.0346)         &    (0.0242)         &    (0.0300)         &    (0.0340)         &    (0.0242)       &    (0.0006)         &    (0.0006)         &    (0.0006)         \\
hukou\_type          &  -0.0421         &     -0.0344         &      0.0332         &     -0.0589\sym{*}  &     -0.0567         &      0.0332       &     -0.0008\sym{*}  &     -0.0007\sym{*}  &     -0.0008\sym{*}  \\
                    &    (0.0350)         &    (0.0442)         &    (0.0240)         &    (0.0355)         &    (0.0447)         &    (0.0240)   &    (0.0004)         &    (0.0004)         &    (0.0004)         \\
health\_status       &    -0.0163         &     -0.0083         &      0.0004         &     -0.0165         &     -0.0056         &      0.0004              &     -0.0002\sym{**} &     -0.0002\sym{*}  &     -0.0002\sym{*}  \\
                    &   (0.0147)         &    (0.0135)         &    (0.0112)         &    (0.0148)         &    (0.0136)         &    (0.0112)    &    (0.0001)         &    (0.0001)         &    (0.0001)         \\
hh\_income &     0.0012         &     -0.0066         &     -0.0148\sym{**} &     -0.0020         &     -0.0076         &     -0.0148\sym{**}     &      0.0001         &      0.0001         &      0.0001         \\
                    &   (0.0099)         &    (0.0080)         &    (0.0075)         &    (0.0099)         &    (0.0083)         &    (0.0075)         &    (0.0001)         &    (0.0001)         &    (0.0001)         \\
age                 &     -0.1944\sym{***}&     -0.1885\sym{***}&     -0.1943\sym{***}&     -0.1930\sym{***}&     -0.1889\sym{***}&     -0.1943\sym{***}&     -0.0024\sym{***}&     -0.0023\sym{***}&     -0.0023\sym{***}\\
                    &    (0.0077)         &    (0.0080)         &    (0.0065)         &    (0.0074)         &    (0.0083)         &    (0.0065)      &    (0.0002)         &    (0.0002)         &    (0.0002)         \\
age$^2$    &        0.0019\sym{***}&      0.0018\sym{***}&      0.0022\sym{***}&      0.0018\sym{***}&      0.0018\sym{***}&      0.0022\sym{***}&      0.0000\sym{***}&      0.0000\sym{***}&      0.0000\sym{***}\\
                    &      (0.0001)         &    (0.0001)         &    (0.0001)         &    (0.0001)         &    (0.0001)         &    (0.0001)      &    (0.0000)         &    (0.0000)         &    (0.0000)         \\
schooling           &      0.0366\sym{***}&      0.0311\sym{***}&      0.0052         &      0.0406\sym{***}&      0.0461\sym{***}&      0.0052    &      0.0003\sym{***}&      0.0002\sym{***}&      0.0002\sym{***}\\
                    &     (0.0051)         &    (0.0061)         &    (0.0039)         &    (0.0052)         &    (0.0070)         &    (0.0039)          &    (0.0000)         &    (0.0000)         &    (0.0000)         \\
pioneer          &     1.7622\sym{***}&      1.5286\sym{***}&      1.1741\sym{***}&      1.8058\sym{***}&      1.5418\sym{***}&      1.1741\sym{***}&                    0.0713\sym{***}&      0.0705\sym{***}&      0.0692\sym{***}       \\&    (0.0657)         &    (0.1024)         &    (0.0415)         &    (0.0670)         &    (0.1093)         &    (0.0415)    &    (0.0042)         &    (0.0041)         &    (0.0042)       \\
distance\_Ingdppc       &      0.8749\sym{***}&      0.5826\sym{***}&      0.2135\sym{***}&      0.8861\sym{***}&      0.5819\sym{***}&      0.2135\sym{***}&          0.0133\sym{***}&      0.0086\sym{**} &      0.0107\sym{**} \\
                    & (0.0888)         &    (0.0934)         &    (0.0376)         &    (0.0881)         &    (0.0948)         &    (0.0376)        &    (0.0032)         &    (0.0041)         &    (0.0045)         \\
distance\_ coop &       -0.0129         &      0.0068         &     -0.0073\sym{*}  &     -0.0121         &      0.0069         &     -0.0073\sym{*}        &     -0.0005\sym{**} &     -0.0004         &     -0.0005         \\
                    &   (0.0082)         &    (0.0144)         &    (0.0039)         &    (0.0082)         &    (0.0143)         &    (0.0039)           &    (0.0003)         &    (0.0003)         &    (0.0003)         \\
distance\_medical    &     46.4214\sym{*}  &     43.7924         &     26.8754\sym{**} &     45.4386         &     45.3497         &     26.8756\sym{**}       &      2.5699\sym{***}&      2.2949\sym{**} &      2.6314\sym{**} \\
                    &    (27.9812)         &   (42.9833)         &   (12.4029)         &   (27.9493)         &   (43.6284)         &   (12.4029)      &    (0.8988)         &    (0.9026)         &    (1.0154)         \\
distance\_highEdu&      3.8617         &      6.5296         &      0.4295         &      3.9861         &      6.7124         &      0.4295        &      0.1378         &      0.2271\sym{**} &      0.2377\sym{**} \\
                    &   (2.4409)         &    (4.4566)         &    (0.7884)         &    (2.4537)         &    (4.4735)         &    (0.7884)       &    (0.0893)         &    (0.0882)         &    (0.0919)         \\
distance\_ppDen       &       0.0464         &      0.1925\sym{**} &     -0.0033         &      0.0475         &      0.1953\sym{**} &     -0.0033     &      0.0046         &      0.0159\sym{***}&      0.0160\sym{***}\\
                    &     (0.0771)         &    (0.0809)         &    (0.0205)         &    (0.0773)         &    (0.0819)         &    (0.0205)           &    (0.0043)         &    (0.0050)         &    (0.0049)         \\
distance\_tertiaryRatio&    0.2462         &      1.8194\sym{**} &     -0.1408         &      0.2182         &      1.7686\sym{**} &     -0.1408  &     -0.0269         &     -0.0673\sym{**} &     -0.0847\sym{**} \\
                    &    (0.5814)         &    (0.7838)         &    (0.2204)         &    (0.5774)         &    (0.7913)         &    (0.2204)   &    (0.0300)         &    (0.0323)         &    (0.0347)         \\
Constant            &     -1.8388\sym{***}&     -1.5829\sym{***}&      0.0083         &     -1.9502\sym{***}&     -1.7668\sym{***}&      0.0083  &      0.0612\sym{***}&      0.0593\sym{***}&      0.0593\sym{***}\\
                    &    (0.3101)         &    (0.2573)         &    (0.2468)         &    (0.3137)         &    (0.2857)         &    (0.2468)     &    (0.0046)         &    (0.0046)         &    (0.0045)         \\
\midrule
Level 2 variance    &    0.1484                &  1.6617            &      1.2686    &     0.1870 &  0.2416      & 0.0000            &                     &                     &                     \\
                    &              (0.0372)         &   (0.1672)       & (0.1010)           &  (0.0312)  &     (0.0455)      &    (0.0000)           &                     &                     &                     \\
Level 3 variance    &                    &                &      0.1054 &   1.6046         & 1.2686           &                   &                     &                     \\ &                   &                     &  (0.0398)   &    (0.1685)       &    (0.1010)        &                     &                     &                     \\
ICC  &            0.0432                &    0.3356         & 0.2783          &   0.0816  &    0.3595     & 0.2783       &                     &                     &                   \\
&            (0.0104)                &    (0.0224)     & (0.0160)              &  (0.0112)   &       (0.0214)    & (0.0160)     &                     &                     &                   \\
\addlinespace
Time FE &            Y         &        Y       &        Y  &        Y        &          Y             &         Y              &         Y              &                    Y   &     Y                 \\
Industry FE &                     &                    &      &   &                &                      &         Y              &                    Y   &     Y                 \\
\addlinespace
Num. of instruments &                     &     &  &                 &                     &                     &    182                &                    180 &     201                \\
AR(2) &                     &                     &       &    &                 &                     &    0.772                &                    0.831 &   0.824               \\
Hansen's J test&                     &      &  &                &                     &                     &      0.368               &                    0.278 &       0.408              \\
\addlinespace
\multicolumn{3}{l}{\textit{Difference-in-Hansen tests}}   \\
\multicolumn{5}{l}{GMM instruments for levels -- Excluding group}                 &           &              & 0.123 & 0.103 & 0.113 \\
\multicolumn{6}{l}{GMM instruments for levels -- Difference (null H = exogenous)}                                           & & 0.466 & 0.377 & 0.622 \\
\multicolumn{6}{l}{GMM instrument for distance\_trend -- Excluding group}        &         &0.553 & 0.381 & 0.521 \\
\multicolumn{7}{l}{GMM instrument for distance\_trend -- Difference (null H = exogenous)}       &0.103 & 0.179 & 0.185 \\
\addlinespace
Obs                 & 749219         & 749219    & 749219  & 749219      & 749219         & 749219         & 729965       & 729965 & 729965         \\
\bottomrule
\insertTableNotes 
\end{longtable}
}
\end{ThreePartTable}

\newpage
\begin{ThreePartTable}
\linespread{0.8}
\begin{TableNotes}[flushleft]\footnotesize
\item Notes: The variable `distance\_JobGrowth' is defined as $GR_{ik,t}-GR_{ij,t}$. Standard errors shown in parentheses are clustered at the destination city. Coefficients are displayed as 0.0000 because the values are smaller than 0.0001. \sym{*} \(p<0.10\), \sym{**} \(p<0.05\), \sym{***} \(p<0.01\)
\item Source: Created by authors using CHFS and China Data Institute (2021).
\end{TableNotes}

\def\sym#1{\ifmmode^{#1}\else\(^{#1}\)\fi}
\fontsize{9}{12}\selectfont{
\setlength{\tabcolsep}{8pt}
\begin{longtable}{l  c c c c c c}
\captionsetup{belowskip=0pt,aboveskip=1pt}
\caption{Determinants of Migration Decisions (1997--2017): Employment Growth Rates\label{tab4}}\\
\toprule\endfirsthead\midrule\endhead\midrule\endfoot\endlastfoot
&\multicolumn{2}{c}{OLS}&\multicolumn{4}{c}{Multilateral Resistance to Migration} \\ \cmidrule(lr){2-3} \cmidrule(lr){4-7} 
                    &\multicolumn{1}{c}{(1)}  &\multicolumn{1}{c}{(2)}  &\multicolumn{1}{c}{(3)}  &\multicolumn{1}{c}{(4)}  &\multicolumn{1}{c}{(5)}  &\multicolumn{1}{c}{(6)}  \\
\midrule
distance\_JobGrowth           &      0.0035\sym{***}&      0.0006         &      0.0002         &      0.0006         &      0.0005         &      0.0001   \\
                    &    (0.0008)         &    (0.0005)         &    (0.0004)         &    (0.0003)         &    (0.0006)         &    (0.0005)        \\
gender            &                    &      0.0009\sym{***}&      0.0006\sym{***}&      0.0001         &      0.0006\sym{***}&      0.0006\sym{***}\\
                    &                     &    (0.0001)         &    (0.0001)         &    (0.0001)         &    (0.0001)         &    (0.0001)       \\
marriage            &                     &     -0.0079\sym{***}&     -0.0074\sym{***}&     -0.0041\sym{***}&     -0.0072\sym{***}&     -0.0074\sym{***}\\
                    &                     &    (0.0006)         &    (0.0005)         &    (0.0005)         &    (0.0005)         &    (0.0005)     \\
hukou\_type          &                     &     -0.0010\sym{**} &     -0.0004\sym{*}  &     -0.0000         &     -0.0004\sym{*}  &     -0.0004\sym{*}      \\
                    &                     &  (0.0004)         &    (0.0003)         &    (0.0002)         &    (0.0003)         &    (0.0003)      \\
health\_status       &                     &     -0.0002\sym{**} &     -0.0001\sym{*}  &     -0.0001\sym{*}  &     -0.0001\sym{*}  &     -0.0001         \\
                    &                     &    (0.0001)         &    (0.0001)         &    (0.0001)         &    (0.0001)         &    (0.0001)         \\
hh\_income &                     &        0.0000         &     -0.0000         &     -0.0001\sym{**} &      0.0000         &     -0.0000         \\
                    &                     &    (0.0001)         &    (0.0001)         &    (0.0001)         &    (0.0001)         &    (0.0001)       \\
age                 &                     &     -0.0024\sym{***}&     -0.0023\sym{***}&     -0.0014\sym{***}&     -0.0023\sym{***}&     -0.0023\sym{***}\\
                    &                     &     (0.0002)         &    (0.0002)         &    (0.0002)         &    (0.0002)         &    (0.0002)          \\
age$^2$    &                     &       0.0000\sym{***}&      0.0000\sym{***}&      0.0000\sym{***}&      0.0000\sym{***}&      0.0000\sym{***}\\
                    &                     &   (0.0000)         &    (0.0000)         &    (0.0000)         &    (0.0000)         &    (0.0000)        \\
schooling           &                     &      0.0002\sym{***}&      0.0002\sym{***}&      0.0000         &      0.0002\sym{***}&      0.0002\sym{***}\\
                    &                     &    (0.0000)         &    (0.0000)         &    (0.0000)         &    (0.0000)         &    (0.0000)         \\
pioneer          &                     &       0.0670\sym{***}&      0.0622\sym{***}&      0.0667\sym{***}&      0.0600\sym{***}&      0.0617\sym{***}\\
                    &                     &    (0.0044)         &    (0.0036)         &    (0.0046)         &    (0.0034)         &    (0.0036)         \\
distance\_Ingdppc       &                     &       0.0141\sym{***}&     -0.0011         &      0.0023         &     -0.0017         &     -0.0031           \\
                    &                     &   (0.0023)         &    (0.0045)         &    (0.0048)         &    (0.0055)         &    (0.0045)       \\
distance\_coop &                     &      -0.0004         &     -0.0007\sym{**} &     -0.0007\sym{**} &     -0.0008\sym{**} &     -0.0009\sym{***}\\
                    &                     &    (0.0003)         &    (0.0003)         &    (0.0003)         &    (0.0004)         &    (0.0003)          \\
distance\_medical    &                     &       2.2484\sym{**} &      3.2036\sym{***}&      2.9437\sym{***}&      4.0403\sym{***}&      3.2219\sym{***}\\
                    &                     &     (0.9989)         &    (0.9130)         &    (0.7792)         &    (1.2303)         &    (1.0067)         \\
distance\_highEdu&                     &       0.1465\sym{*}  &      0.1033         &      0.0485         &      0.1771\sym{*}  &      0.1322     \\
                    &                     &    (0.0815)         &    (0.0940)         &    (0.0885)         &    (0.0964)         &    (0.0957)        \\
distance\_ppDen       &                     &     0.0025         &     -0.0020         &      0.0005         &     -0.0099         &     -0.0008        \\
                    &                     &   (0.0019)         &    (0.0083)         &    (0.0073)         &    (0.0075)         &    (0.0079)    \\
\multicolumn{2}{l}{distance\_tertiaryRatio}                    &       0.0034         &     -0.0374         &     -0.0410         &     -0.0291         &     -0.0365        \\
                    &                     &    (0.0188)         &    (0.0264)         &    (0.0261)         &    (0.0272)         &    (0.0264)      \\
Constant            &      0.0088\sym{***}&      0.0616\sym{***}&      0.0600\sym{***}&      0.0394\sym{***}&      0.0601\sym{***}&      0.0597\sym{***}\\
                    &    (0.0009)         &    (0.0045)         &    (0.0042)         &    (0.0042)         &    (0.0042)         &    (0.0042)         \\
\midrule
Time FE & Y   & Y  & Y & Y &  &  \\
Industry FE &   &  & Y   & Y & Y & Y \\
Origin FE &   &   & Y   &  &  Y & \\
Destination  FE &   &   & Y   &  & & Y \\
Pairs of cities FE &  &  &   & Y & & \\
Origin-year FE & & &  & & & Y \\
Dest-year FE & &  & & & Y & \\
$R^2$ & 0.0016       & 0.0458        & 0.0534            & 0.0840         & 0.0722  & 0.0698       \\	
Obs                 & 1038407         & 758754         & 739308         & 739303         & 739109         & 739067        \\
\bottomrule
\insertTableNotes 
\end{longtable}
}
\end{ThreePartTable}

\newpage
\section*{Appendix B: Figures and Statistics}
\setcounter{table}{0}
\setcounter{figure}{0}
\renewcommand{\thetable}{B\arabic{table}}
\renewcommand{\thefigure}{B\arabic{figure}}

\begin{figure}[!htb]	
\centering
\captionsetup{justification=centering}	
\includegraphics[height=3.5in,width=5.2in]{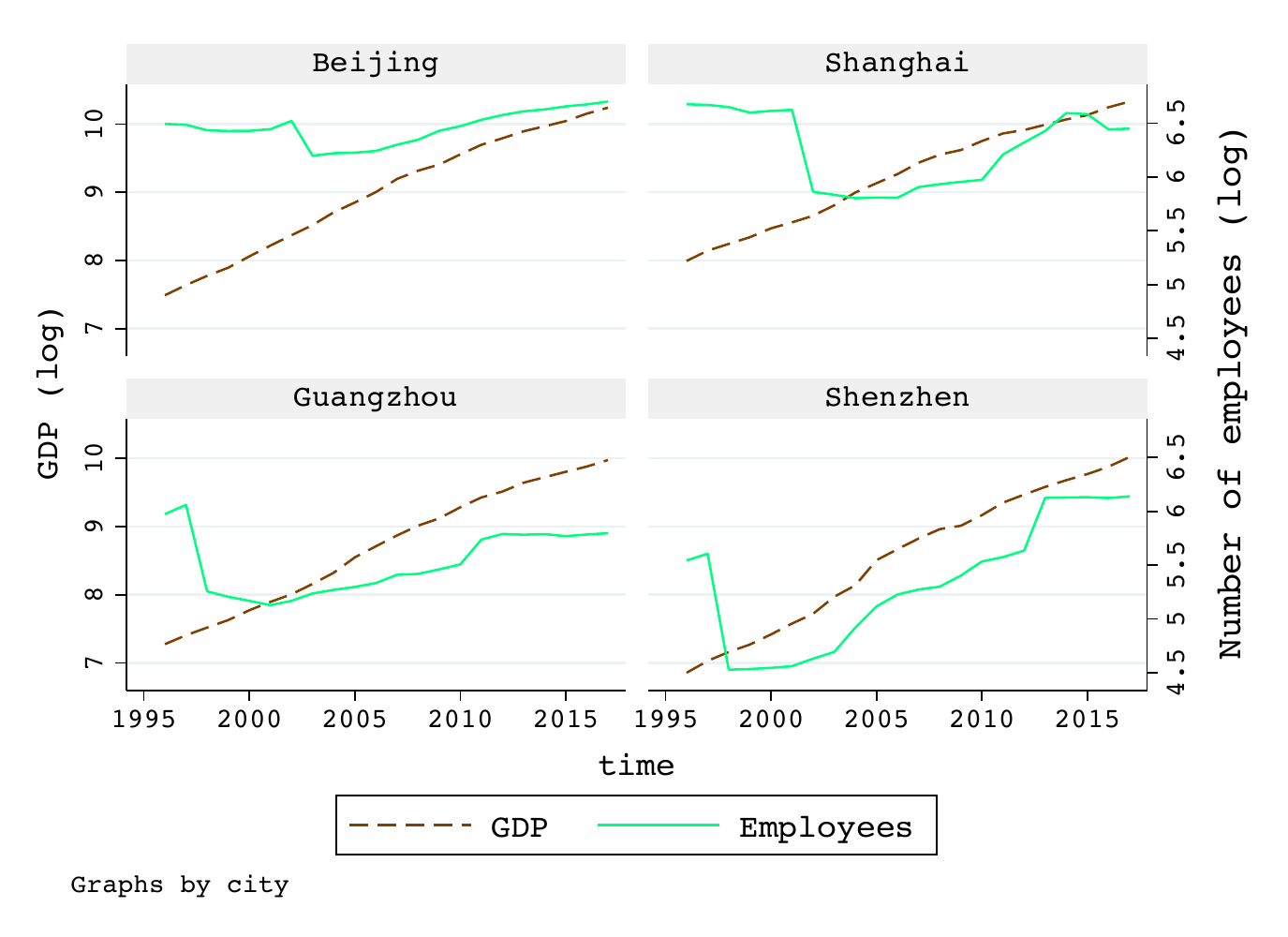}
\caption{\fontsize{10}{12}\selectfont{GDP growth vs. employment levels in `Bei-Shang-Guang-Shen'. \\Source: Author's elaboration using China Data Institute (2021).}\label{figb2}}
\end{figure}

\begin{figure}[!htb]	
\centering
\captionsetup{justification=centering}	
\includegraphics[height=3.5in,width=5.2in]{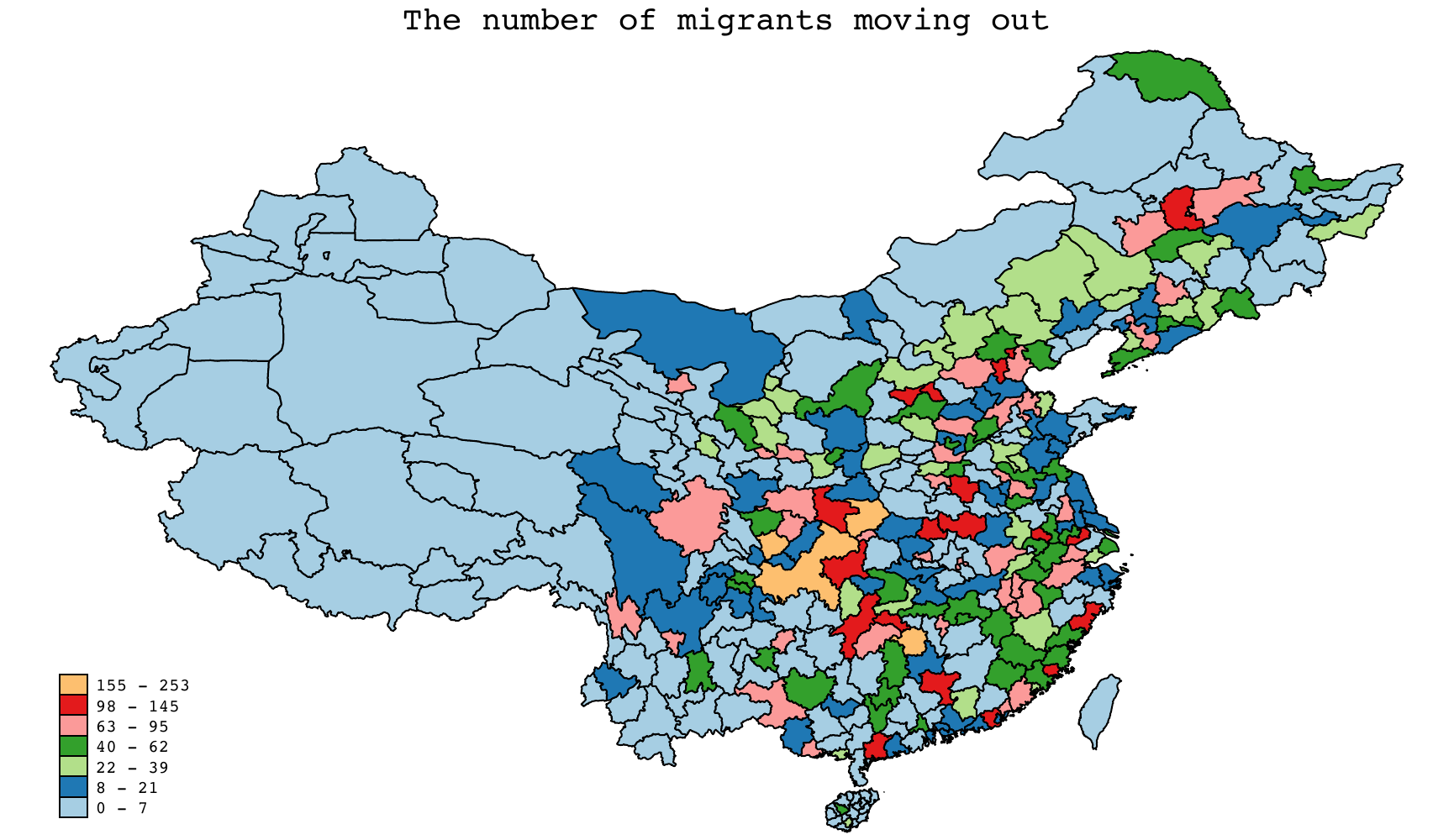}
\caption{\fontsize{10}{12}\selectfont{Geographic distribution of emigrants across Chinese prefecture-level cities. \\Source: Author's elaboration using CHFS.}\label{figb1}}
\end{figure}

\newpage
\section*{Appendix C: Data Wrangling Report\label{tab9}}
\setcounter{table}{0}
\setcounter{figure}{0}
\setcounter{footnote}{0}
\renewcommand{\thetable}{C\arabic{table}}
\renewcommand{\thefigure}{C\arabic{figure}}
\fontsize{10}{12}\selectfont{

\subsection*{Introduction}
\hspace{1.4em}The China Household Financial Survey was conducted biannually between 2011 and 2017. The latest wave tracked the majority of preceding households plus newly-surveyed families where totally 127,012 individuals are involved.\footnote{Some households were surveyed previously but lost in later waves.} Each wave has three sets of data providing information on individuals, households, and cities. We used the latest wave because the information on individuals' origin cities are absent in earlier waves, and it satisfies all the features our approach needs: (a) it includes both natives and migrants, and (b) enables us to compile origin information at the prefecture city level.

\hspace{1.4em}It should be noticed that, as a household survey, answers with regard to family members were made by the respondent who acted as a delegate of his or her family. The CHFS's interviewers chose the person who was most familiar with his or her household's economic conditions, though this might not be feasible sometimes. In other words, the respondent can either be him- or herself or share certain family relationships with the persons who were questioned (parent, spouse, etc.). This design entails a few concerns. For example, when the respondents were on their own behalf, their resident cities are not clearly stated in the data since the question only asked where their family members were living now, if they did not live together.  Instead, we can learn in which city respondents got surveyed from the city table. Yet where people got surveyed are not always the city they mostly lived in. In two ways, we clearly identified this discrepancy. Literally, as the 2017 questionnaire asked if the current city/county was the place where the family's main economic activities were carried out, an individual who did not stay with family members but visited or went back when the survey was conducted can be a respondent, although intuitively they should not be. While technically, two variables help us check if individuals lived in their Hukou registration city/county/town. Based on them, we can make a comparison between the Hukou registration city and surveyed city. A few thousands of observations are found dubious. For instance, the variable A2019b shows that the individual lived in his or her Hukou registration city but he or she was surveyed in another city. To clean and compile the data, when surveyed cities are used as resident cities, we only retained householder samples.  

\subsection*{Migrant Identification}
\begin{figure}[h]
\centering	
\captionsetup{justification=centering}	
\includegraphics[scale=0.3]{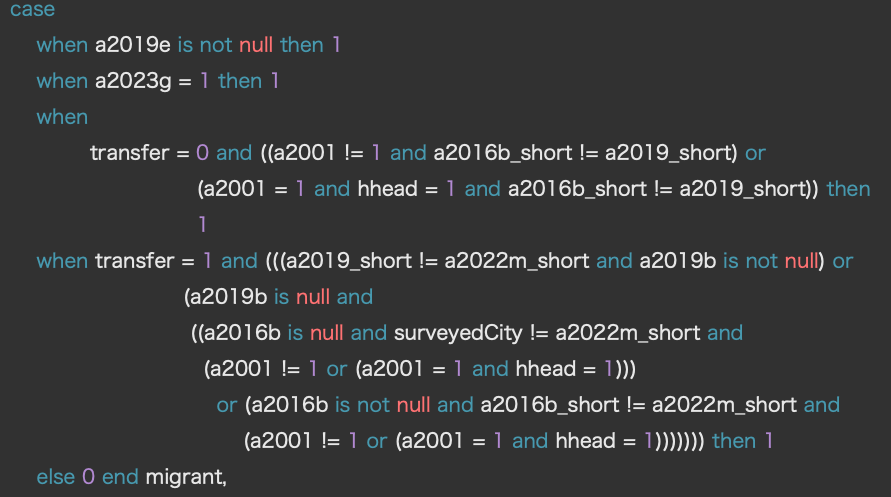}
\caption{\fontsize{10}{12}\selectfont{SQL codes for identifying migrants.}}
\end{figure}

\hspace{1.4em}Overall, there are four types of migrants. First, based on the variable A2019e, we distinguished the floating population from natives and migrants with Hukou transfer.\footnote{This question asked `in which year did he/she leave the city where his/her Hukou is registered'. All the variables discussed in this report are summarised in \hyperref[tab10]{Table C1}.}  The 2017 questionnaire annotates that the referring question would only be delivered if individuals' Hukou registration cities are different from their resident cities. As shown in Figure C1, we specified A2019e to be not null. Also, sometimes the data is messy, displaying information dissimilar to what we are told by the questionnaire, as shown in Figure C2. To avoid mistreating certain migrants as natives resulting from missing inputs in A2019e, we coded a supplementary condition starting by transfer=0.\footnote{Variables marked with `short' indicate their inputs are based on the first four digits of the NBS codes. This is to ensure that migrants are identified at the prefecture level.} Another type that we can straightforwardly identify is returnees. When the CHFS system detected that the individual's Hukou registration city is identical to his/her resident city, the interviewer then moved on to the question A2023g -- if he/she had ever left the Hukou registration city for somewhere else to for more than six months. It is worth noting, although interviewers only asked family members who were aged above 16 in 2017, returnees can be quite young when they migrated. In our final dataset, these observations were dropped.

\begin{figure}[h]	
\centering
\captionsetup{justification=centering}	
\includegraphics[scale=0.4]{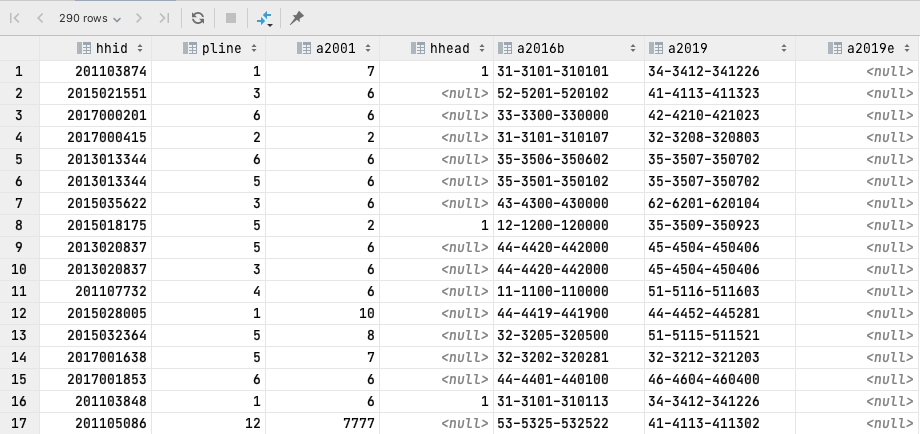}
\caption{\fontsize{10}{12}\selectfont{Respondents whose Hukou registration cities and resident cities are mismatching. \\ Notes: These are failed to be identified by A2019e.}}
\end{figure}

\hspace{1.4em} Identifying another two types is more demanding. By looking at the variable A2019e, we can learn whether the individual was living in the Hukou registration city or not, but there were some migrants with Hukou transfer. To avoid overestimating natives, we need other variables to help us find these samples, namely, A2022k (called `transfer' in codes) and A2022m. The codes perhaps look a bit lengthy but basically, we found four cases --  migrant who lived in the county where his/her new Hukou was registered; migrant who lived outside the Hukou registration county but inside the newly-registered prefecture city; migrant who lived outside the newly-registered prefecture city; migrant whose Hukous had been transferred to another prefecture city but remained staying in his/her origin cities (as shown in Figure C3). To exclude samples of the last case that are not captured by A2019e, we used surveyed city followed by A2001.\footnote{Recall that A2001 describes family relationships between respondents and interviewed persons.} Lastly, a few people had migrated for several times. The situation can be rather intricate -- they transferred the Hukous after they migrated to new prefecture cities, yet the cities where their Hukous were newly registered are not the ones paired with the variables telling in which year they migrated. We specified these samples as migrants first and then dropped some of them if, after compiling, their destination cities and moving-in years are still mismatching.\footnote{We discussed how we managed it in detail in the next section.}

\begin{figure}[t]
\centering		
\captionsetup{justification=centering}
\includegraphics[scale=0.3]{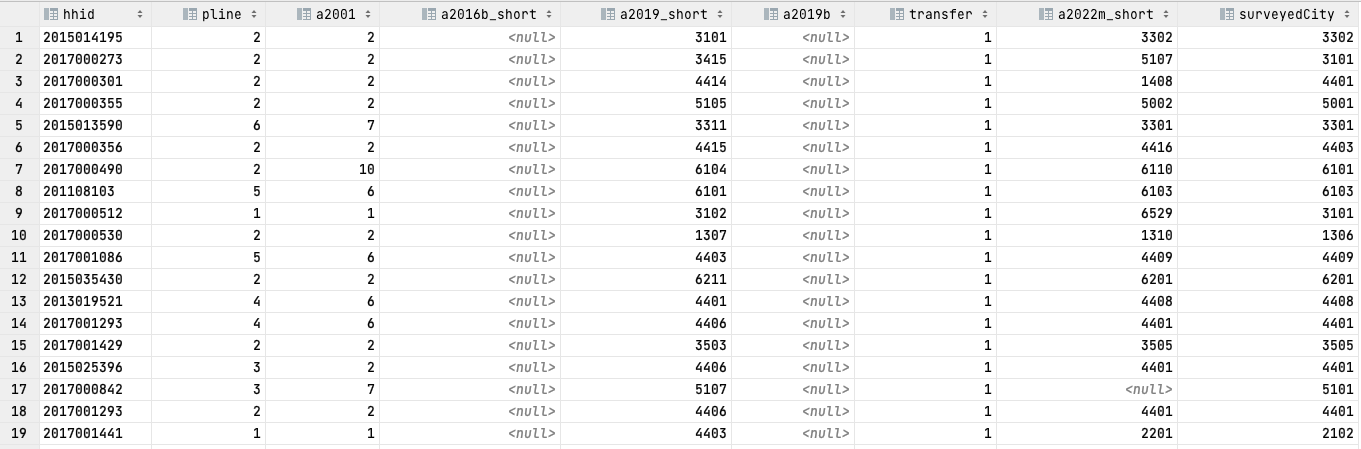}
\caption{\fontsize{10}{12}\selectfont{Migrants with Hukou transfer: Surveyed cities are identical to origin cities.}}
\end{figure}

\begin{figure}[h]
\centering		
\captionsetup{justification=centering}
\includegraphics[scale=0.3]{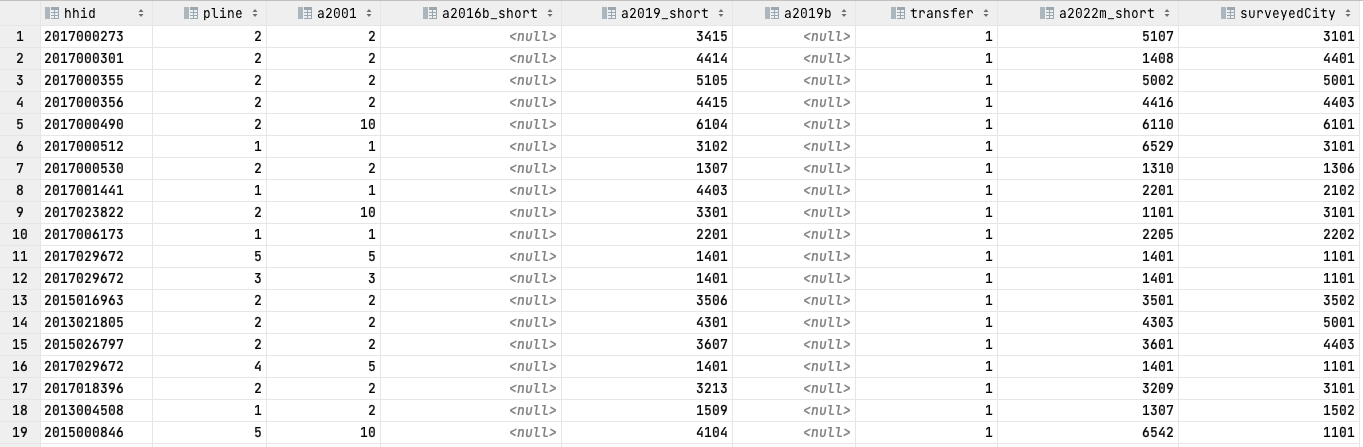}
\caption{\fontsize{10}{12}\selectfont{Migrants with Hukou transfer: New and old Hukou registration cities as well as surveyed city are all non-identical.}}
\end{figure}

\subsection*{Information Pairwise}
\hspace{1.4em}Now we successfully sorted out migrant observations, while the next issue lying ahead of us is to identify to which prefecture city that each migrant moved and accordingly, in which year they moved out and moved in. Foremost, variables A2016b, A2019, A2023j and `surveyed city' all provide usable information. To apply the correct one, we need to further classify migrants into different situations. As always, we endeavoured to improve the data accuracy and retain as many observations as possible.

\begin{figure}[h]
\centering		
\captionsetup{justification=centering}
\includegraphics[scale=0.3]{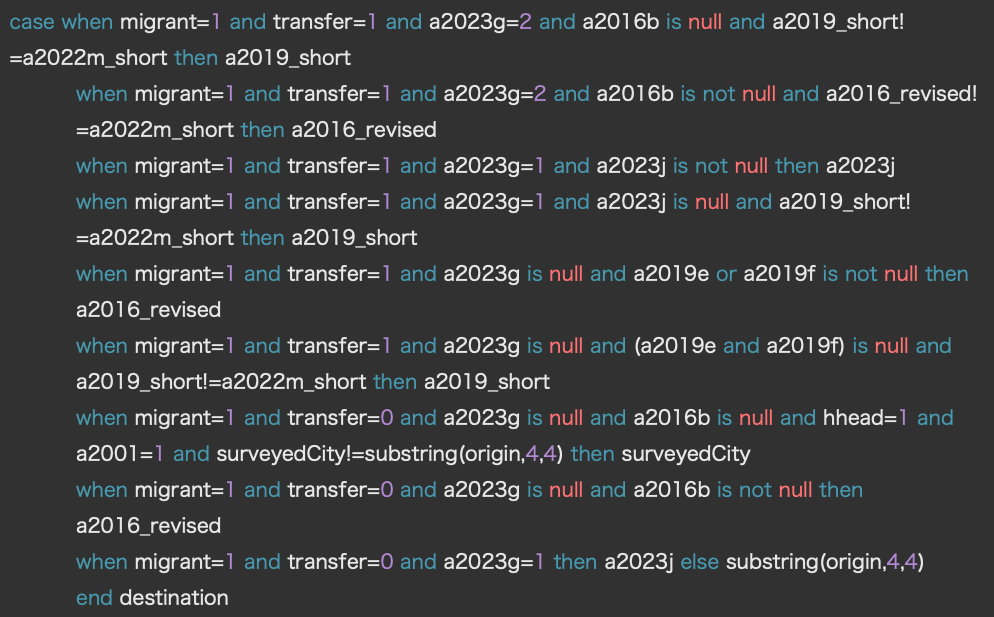}
\caption{\fontsize{10}{12}\selectfont{SQL codes for identifying migration destination.}}
\end{figure}

\hspace{1.4em}Generally, the classification can be summarised as follows:

\hspace{1.4em}a. When A2023g equals 2, it suggests that individuals had never resided in other prefecture cities. While looking at migrants with Hukou transfer, sometimes the reference point that respondents took is the cities where their current Hukous were registered, instead of their `real' origin cities. Therefore, by setting transfer equals 1 and A2023g equals 2 followed by A2016b, we assigned either current Hukou registration cities or resident cities to those samples as their destination in terms of their specific situations.

\hspace{1.4em}b. When A2023g equals 1, it simply means that individuals were returnees. In general, A2023j tells us where individuals had migrated to before they returned to the Hukou registration cities. Yet for certain observations, this information is missing. Instead, we assigned current Hukou registration cities as the destination to some of them if they are migrants with Hukou transfer.

\hspace{1.4em}c. When A2023g is null, the CHFS annotates that this question was not assigned to those individuals because they lived in the prefecture cities that their Hukous did not pertain to. In this case, we could learn about their resident cities according to A2016b. However, sometimes even though we know in which city the migrant was living, his/her moving-out and/or moving-in year, referring to A2019e and A2019f, is still missing and thus, these observations cannot be used. As an alternative solution to keeping information pairwise, we assigned the new Hukou registration cities to them and re-calculated their migration years by assuming they moved one year before the time they transferred the Hukous.\footnote{In large cities, obtaining a local Hukou is quite hard for immigrants especially since 2014, so it may take a much longer period for people to successfully transfer their Hukous. However, the number of such observations after cleaning is just 745 and among them, migrants who transferred their Hukous to Tier-1 cities are very few. We also confirm that the statistical significance of our trending indicator holds in the case of excluding those observations.}  

\hspace{1.4em}d. Lastly, among the floating population, destination cities can either be A2016b or `surveyed city' depending on if A2016 has a factual value. Again, we restricted the use of the variable `surveyed city' to householders (hhead=1).

\hspace{1.4em}Besides the aforementioned, we assigned the origin cities (A2022m) to all natives as their destination. Moreover, as shown in Figure C6, sometimes the moving-out year (A2019e) is not the same as the moving-in year (A2019f) because individuals moved more than once. As they are not useful in our analysis, we dropped these observations.

\begin{figure}[h]
\centering		
\captionsetup{justification=centering}
\includegraphics[scale=0.35]{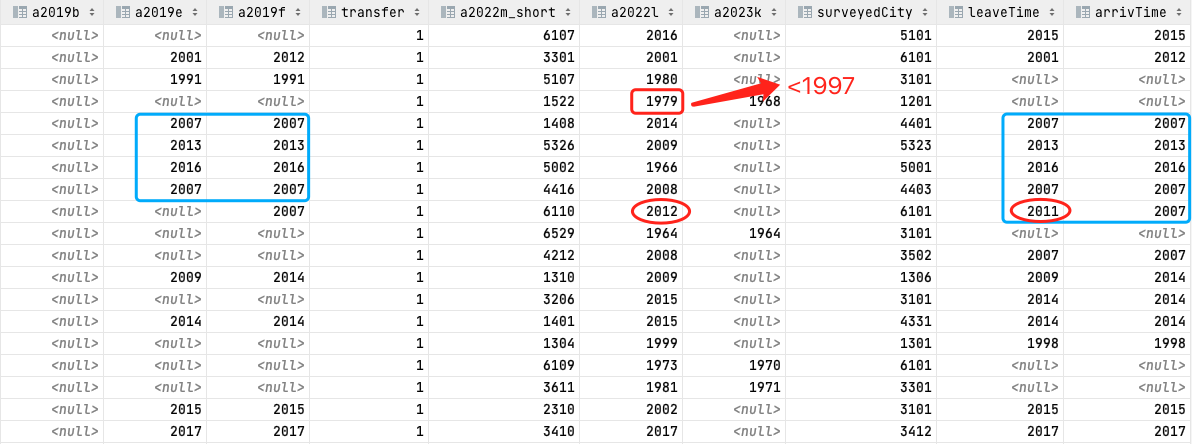}
\caption{\fontsize{10}{12}\selectfont{Example of migration year.}}
\end{figure}

\subsection*{Final Cleaning Procedure}
\paragraph{Migrant Samples} \mbox{}

\hspace{1.4em}Afterwards, there is still a bit of information not pleasing us because some observations' inputs remained weird. For instance, the destination city does not necessarily match the moving-in year, and a small cohort of respondents did not answer questions based on the common knowledge, as shown in Figure C7. First, dozens of samples present available inputs in A2023j, but the corresponding data that should be stored in A2023k is missing. Hence, the destination variable extracted the data from A2023j but used the proxy (the year of Hukou transfer minus 1) as the migration time. To fix this problem, we can either replace the destination variable with A2019 (the current Hukou registration city) or simply drop them. We chose the latter. Moreover, some observations have identical inputs in A2022m and A2023j. This could be explained by three possible reasons -- the respondent took his/her current Hukou registration city as the reference point and thus, considered where he/she came from as the place he/she had stayed in outside the Hukou registration city; individuals migrated and transferred the Hukous more than once; individuals returned to their origin cities even after they transferred the Hukous. By setting A2022l$<$A2023k, we can easily identify the last group, but it is hard to distinguish the first two. As a result, although we know they are migrants, we cannot identify in which year these migrants moved to their resident cities. Thus, these observations were eventually dropped as well.

\begin{figure}[h]
\centering		
\captionsetup{justification=centering}
\includegraphics[scale=0.3]{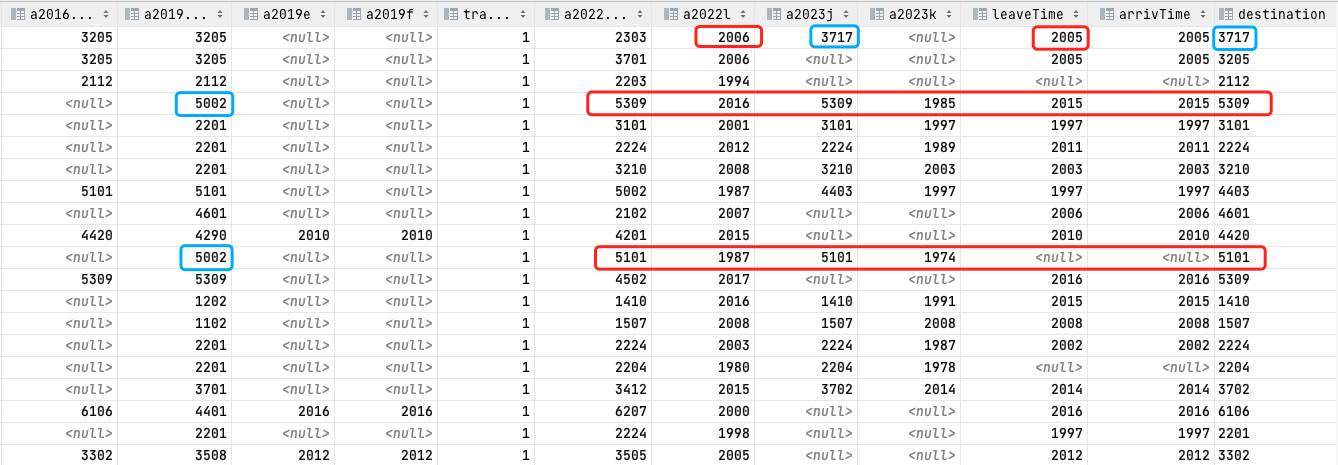}
\caption{Remaining errors.}
\end{figure}

\hspace{1.4em}Recall that certain returnees may migrate when they were quite young (under 16 years old). As we aim to examine labour migration, teenagers are not proper to be included. After merging the datasets, we transformed the age variable to be time-variant. In this way, we eliminated all the observations if their ages were either below 16 or above 65 between 1997 and 2017. In addition, for migrants who worked previously but did not have a job when they were interviewed in 2017, we matched regional employment statistics for them based on the industry categories of their last employment. Concerning they may migrate after being unemployed or retired, only respondents whose last jobs' termination dates (A3139) are later than their migration time were kept in the data.

\paragraph{Native Samples} \mbox{}

\hspace{1.4em}Native samples are much simpler to clean. Hereby, we give a brief summary:

\hspace{1.4em}$\bullet$ Drop observations if A2022k (transfer) equals 1 but A2019 is null.

\hspace{1.4em}$\bullet$ Drop observations if A2022k (transfer) equals 1 but A2019b and A2023g are null.

\hspace{1.4em}$\bullet$ Drop observations if the city retrieved from the variable `surveyed city' is non-identical to the origin city while the respondent was identified as a householder. 

\hspace{1.4em}Lastly, among the full sample (both natives and migrants), observations were dropped if A3138 equals 2 (who had never worked).  
}

\subsection*{Variables Summary}
\begin{table}[htp]
\fontsize{10}{12}\selectfont{
\captionsetup{belowskip=0pt,aboveskip=1pt}
\noindent\makebox[\textwidth]{
\centering
\begin{threeparttable}
\caption{Variables -- Questions\label{tab10}}
\begin{tabular}{p{3cm} l}    \toprule
\rowcolor{green! 40}	\emph{Variables} & \emph{\hspace{12em}Questions} \\\midrule
A2001 & \uline{XXX} is your? [Example: Myself; Spouse; Parents; Children]\\\addlinespace
A2016b & In which province/city/county does \uline{XXX} live? \\\addlinespace
A2019 & Which province/city/county is the registered residence of \uline{XXX}? \\\addlinespace
A2019b\dag & Is \uline{XXX} Hukou registered in the villages/towns where he/she now lives? \\\addlinespace
A2019e\ddag & In which year did \uline{XXX} leave \uline{[A2019]}? \\\addlinespace
A2019f\ddag & In which year did \uline{XXX} come to his/her resident province/city? \\\addlinespace
A2022k & Has  \uline{XXX} ever transferred the Hukou to another district/county? \\\addlinespace
A2022l & In which year did \uline{XXX} experience his/her latest Hukou transfer?\\\addlinespace
A2022m & \uline{XXX}'s Hukou is moved out from which county/city/province? \\\addlinespace
A2023g$\ast$ & Has \uline{XXX} ever left \uline{[A2019]} for somewhere else to for over 6 months? \\\addlinespace
A2023j$\ast$ & Which province/city did \uline{XXX} live before returning? \\\addlinespace
A2023k$\ast$ & In which year did \uline{XXX} go to \uline{[A2023j]}? \\\addlinespace
A3138 & Has \uline{XXX} worked before? \\\addlinespace
A3139 & When did \uline{XXX}'s last job end? \\\bottomrule
\end{tabular}
\footnotesize{Notes: \dag\ denotes questions that are asked only if the CAPI system detects that the individual's resident county/district corresponds to his/her Hukou registration county/district. \ddag\ denotes questions that are asked only if the CAPI system detects that the individual's resident prefecture city is non-identical to his/her Hukou registration city. $\ast$ denotes questions that are asked only if the CAPI system detects that the resident prefecture city of an individuals whose age had been above 16 by the time of survey is identical to his/her current Hukou registration city. \\ Source: China Household Finance Survey (2017).}\\
\end{threeparttable}
}}
\end{table}

\end{document}